\documentclass[twocolumn]{sig-alternate}
\makeatletter
\def\@copyrightspace{\relax}
\makeatother

\usepackage{dblfloatfix}
\usepackage[table]{xcolor}
\usepackage[american]{babel}
\usepackage{booktabs}
\usepackage{graphicx}
\usepackage{tikz}
\usepackage{color}
\usepackage{tabularx}
\usepackage{varwidth} 
\usepackage{listings}
\usepackage{url}
\usepackage{balance}
\usepackage{amsmath}
\usepackage{enumitem}
\usepackage{caption}
\usepackage{multicol}

\usepackage{xcolor}
\usepackage{framed}
\usepackage{amssymb,amsmath}
\usepackage{fancyvrb}
\usepackage{comment}

\newcommand{\reportnumber}{Report Number: FERMILAB-PUB-25-0356-ETD}

\usepackage[angle=0,scale=1,color=black,firstpage=true,opacity=1]{background}

\SetBgContents{\small\texttt{\reportnumber}}
\SetBgPosition{current page.south east}
\SetBgHshift{5cm}
\SetBgVshift{1cm}

\begin{document}

\title{\LARGE\bf Fast Machine Learning for Quantum Control\\of Microwave Qudits on Edge Hardware}

\numberofauthors{5}
\author{
\alignauthor
Flor Sanders\\
\email{\emph{fps2116@columbia.edu}}
\alignauthor
Gaurav Agarwal\\
\email{\emph{gsa2131@columbia.edu}}
\alignauthor
Luca Carloni\\
\email{\emph{luca@cs.columbia.edu}}
\and
\alignauthor
Giuseppe Di Guglielmo\\
\email{\emph{gdg@fnal.gov}}
\alignauthor
Andy C. Y. Li\\
\email{\emph{cli@fnal.gov}}
\alignauthor
Gabriel N. Perdue\\
\email{\emph{perdue@fnal.gov}}
}


\maketitle


\begin{abstract}
{\small\em
Quantum optimal control is a promising approach to improve the accuracy of quantum gates, but it relies on complex algorithms to determine the best control settings. 
CPU or GPU-based approaches often have delays that are too long to be applied in practice.
It is paramount to have systems with extremely low delays to quickly and with high fidelity adjust quantum hardware settings, where fidelity is defined as overlap with a target quantum state.
Here, we utilize machine learning (ML) models to determine control-pulse parameters for preparing Selective Number-dependent Arbitrary Phase (SNAP) gates in microwave cavity qudits, which are multi-level quantum systems that serve as elementary computation units for quantum computing.
The methodology involves data generation using classical optimization techniques, ML model development, design space exploration, and quantization for hardware implementation.
Our results demonstrate the efficacy of the proposed approach, with optimized models achieving low gate trace infidelity near $10^{-3}$ and efficient utilization of programmable logic resources.
}
\end{abstract}

\section{Introduction}
\label{sec:intro}

Quantum Computing (QC) has undergone rapid development over the past four decades \cite{quantum_computing_review}. 
Harnessing the quantum mechanical properties of the natural world, specialized algorithms such as Shor's Algorithm~\cite{shors_algorithm} and Grover's Algorithm~\cite{grovers_algorithm} theoretically offer asymptotic scaling advantages over all known relevant classical algorithms.

Although the potential of quantum computation is promising, current systems are encumbered by problems relating to the stability of the qubits and qudits, the most elementary QC units~\cite{quantum_computing_limitations}.
The efficient manipulation of quantum systems, better known as quantum control, is an active field of research.
The core idea is that more precise manipulation of the quantum state of the QC will enable better performance given identical hardware~\cite{qudit_quantum_control}.

A qudit is the QC equivalent of an N-ary digit (where $N \geq 2$), whose multilevel and corresponding larger state space allows for the reduction in complexity when implementing quantum algorithms.
They have emerged as an alternative to binary qubits~\cite{qudits_basics,alam2022quantumcomputinghardwarehep}.
Of particular relevance to this work, the generation of such multilevel quantum states in three-dimensional microwave cavities has proven to be a promising paradigm for the implementation of qudit devices~\cite{quantum_rf_cavity_control_gate}.

\subsection{Motivation}

Although mathematical models for quantum control of microwave cavity qudits have been developed, the computational requirements of the classical optimization involved in these approaches hinder their application to real-time systems~\cite{quantum_rf_cavity_control_gate}.
The application of such control models to actual QC hardware requires the development of custom and programmable, low-latency hardware.
An additional consideration for the systems considered here is the need to operate at cryogenic temperatures.
Within these constraints, it is important to keep the computational complexity of the control model low while still maintaining the fidelity of the model.
To this end, the use of Machine Learning (ML) to develop lightweight algorithms that approximate optimal quantum control schemes is an interesting route to consider~\cite{ml4quantum,bhat2024machine}. 

Previous work investigated such a methodology for a quantum gate that rotates a qubit along a particular axis~\cite{ml4quantum,bhat2024machine}.
Their experiments demonstrate that high levels of model fidelity can be achieved with ML models that can comfortably be implemented on a commodity Field Programmable Gate Array (FPGA).
Here, we will further investigate the use of ML models for quantum control applications by developing a new model that enables the control of microwave cavity qudits.

\subsection{Contribution}

We want to create a target action \(U\), depending on a set of input parameters ($\alpha_0$, $\alpha_1$, $\alpha_2$, ...), on the N-state qudit which changes the quantum state \(\ |A\rangle \) into \(U |A\rangle\).
The transformation is achieved using microwave pulses that are capacitively coupled to the quantum device. 
Numerical simulation tools such as Juqbox~\cite{juqbox1,juqbox2}, Quandary~\cite{quandary1,quandary2}, and QuTiP~\cite{Qutip2012}, use techniques such as gradient ascent to compute the optimal pulse parameters ($\theta_0$, $\theta_1$, $\theta_2$, etc.).
Calculating these parameters usually incurs significant latency when these algorithms are executed as software. Furthermore, the power budget in a cryogenic environment is very limited and does not suit the integration of CPUs or GPUs~\cite{Rodriguez-Borbon2024,HybridSimulation2024}.

\begin{figure}[t!]
\centering
\captionsetup{justification=centering, format=hang}
\includegraphics[width=0.85\columnwidth]{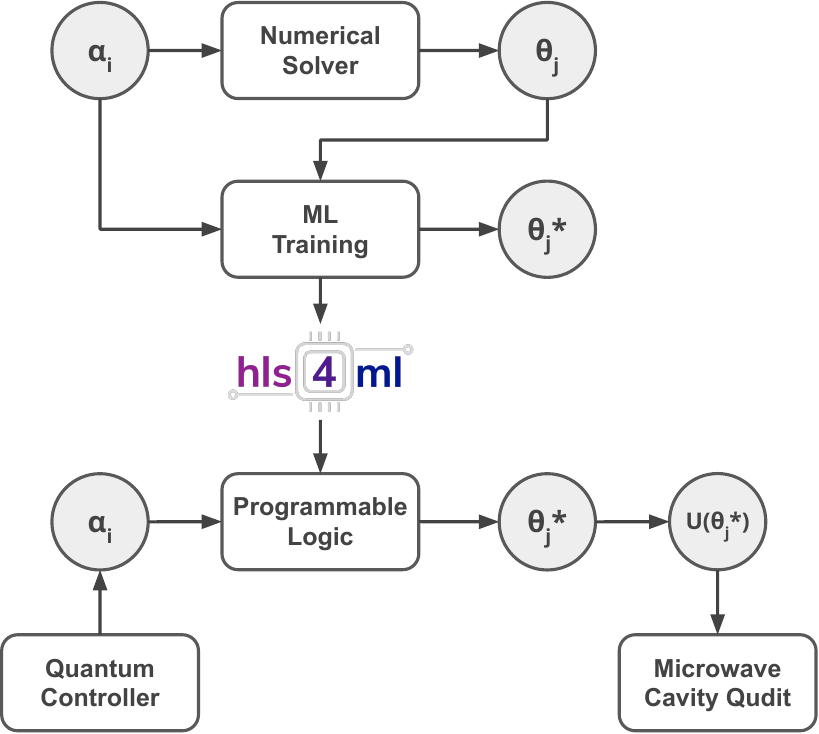}
\caption{System Design Flow}
\label{fig:architecture}
\end{figure}

Hence, this work will focus on developing a deep neural network to be deployed on programmable logic.
The network will predict the optimal pulse configuration parameters ($\theta_0^*$, $\theta_1^*$, $\theta_2^*$, etc.), to produce the time-varying microwave pulses that alter the quantum state of the qudit.
An overview of the system design flow is provided in Fig.~\ref{fig:architecture}.

We trained our neural network in TensorFlow~\cite{tensorflow2015-whitepaper} and translated it into a functional hardware specification in C++ with \texttt{hls4ml}~\cite{Duarte:2018ite,hls4ml2021}.
The C++ specification is input for high-level synthesis (HLS) tools that translate the specification into a hardware implementation at the register-transfer level (RTL), typically written in hardware description languages such as Verilog or VHDL. 
We will use the flexibility of \texttt{hls4ml} combined with the HLS tuning capabilities to optimize the design of a hardware implementation for programmable devices~\cite{siemens2024catapult}. We will compute performance indicators such as model fidelity, latency, and resource utilization as programmable logic.
Our core contribution then is a high-fidelity ML model compiled to a programmable device for controlling a qudit encoded in a three-dimensional resonator.

\section{Methodology}\label{sec:methodology}
 
This section provides a more detailed explanation of the various steps of our design methodology shown in Fig.~\ref{fig:architecture}.
In Sec.~\ref{sec:data} we explain how we produced the dataset used to train the ML model for qudit-based quantum control.
Next, Sec.~\ref{sub:ml-development} discusses the steps required to design and optimize a ML model for this task.
In Sec.~\ref{sec:quantization}, we discuss the quantization process, which is an intermediate step for translating the model to hardware. Quantization optimizes the model for hardware deployment by reducing computational precision while maintaining acceptable accuracy. Subsequently, Sec.~\ref{sec:hls} details the hardware translation process.
 
\subsection{Data Generation}
\label{sec:data}
The microwave pulse function to be approximated implements a Selective Number-dependent Arbitrary Phase (SNAP) gate \cite{Heeres_2015}, which phase shifts qudit level $n$ by a phase angle $\alpha$ and allows universal control of the qudits together with displacement gates \cite{Krastanov_2015}.
The SNAP gate can be represented by a $d$-by-$d$ diagonal matrix 
\begin{equation}
    \mathrm{SNAP}(\alpha, n) = \mathrm{Diag}([e^{j \alpha} \text{  if  } k=n \text{  else  } 1]_{k=1, \cdots, d}).
\end{equation}

In most implementations, the microwave cavity qudit is dispersively coupled to an ancilla qubit to allow nonlinear control \cite{alam2022quantumcomputinghardwarehep,quantum_rf_cavity_control_gate}. In the rotating frame, the qudit-qubit system can be described by the system Hamiltonian $H_0$ with dispersive approximation such that
\begin{equation}
H_0 = -\chi a^{\dagger} a b^{\dagger} b  - \xi (b^{\dagger})^2 b^2,
\end{equation}
where $a$ ($b$) is the annihilation operator of the qudit (qubit),  $\chi$ is the dispersive coupling strength, and $\xi$ is the qubit nonlinearity. Throughout this work, we will take $\chi=5$ MHz and $\xi = 200$ MHz, similar to the values used in reference \cite{quantum_rf_cavity_control_gate}, and truncate the qubit up to three levels.

\begin{figure}
    \includegraphics[width=1.0\columnwidth]{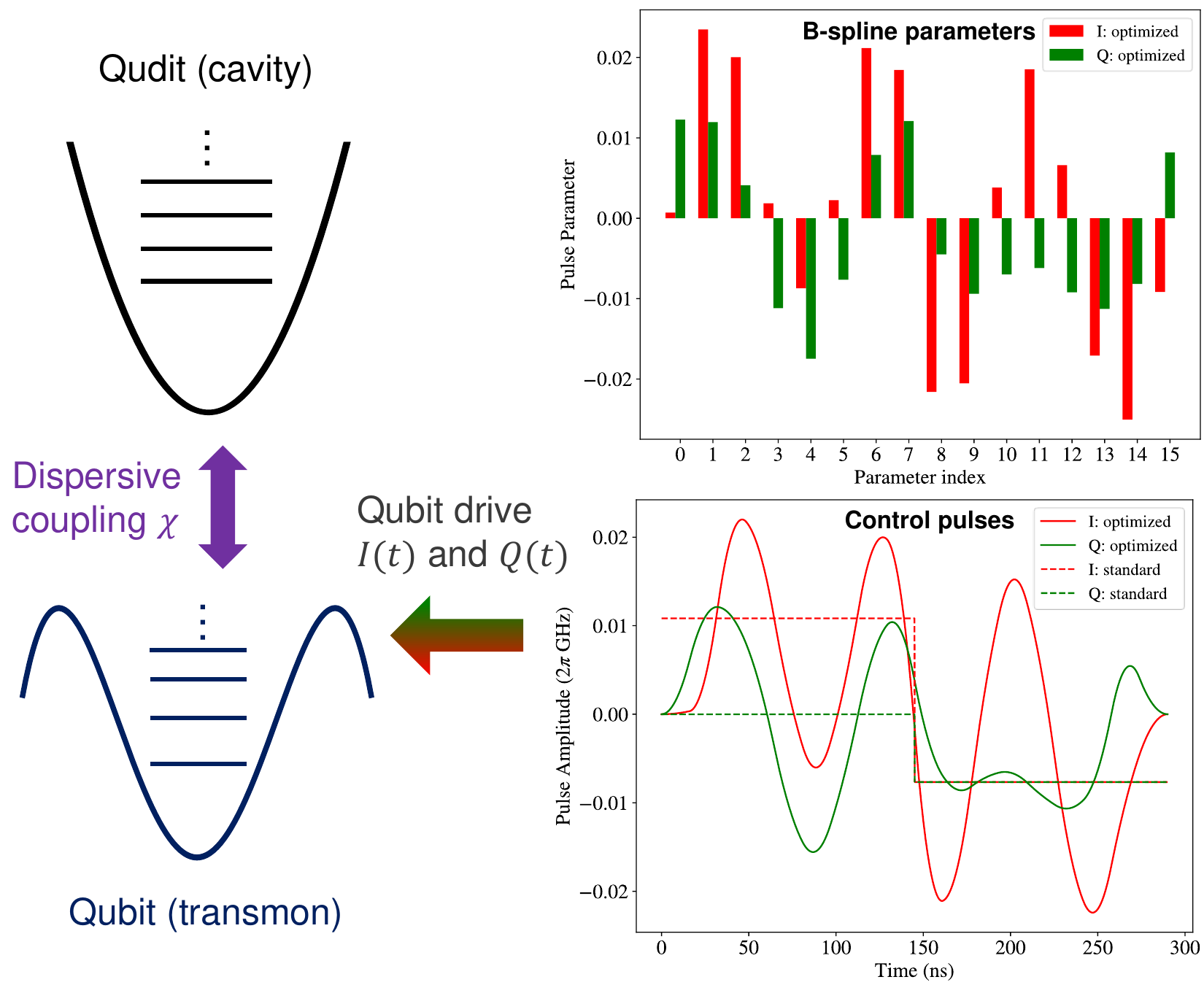}
    \caption{Schematic diagram of the optimal pulse controlling the qudit-qubit system.}
    \label{fig:pulse_gen_schematic}
\end{figure}

To realize the SNAP gate, a microwave control is capacitively coupled to the ancilla qubit, and the control pulse is represented by an in-phase part $I(t)$ and a quadrature part $Q(t)$.
The full Hamiltonian $H$ is given by
\begin{equation}
    H = H_0 + H_{\mathrm{ctrl}}
\end{equation}
with the control Hamiltonian $H_{\mathrm{ctrl}}$,
\begin{equation}
    H_{\mathrm{ctrl}} = I(t)(b + b^\dagger) + j Q(t)(b - b^\dagger).
\end{equation}
In-phase and quadrature pulses $I(t)$ and $Q(t)$ can be parameterized by quadratic B-splines~\cite{petersson2022optimal}. In this work, we take 16 B-spline parameters, each to represent $I(t)$ and $Q(t)$, respectively. These parameters are optimized to minimize the infidelity (the cost function) given by $1 - f$, where $f$ is the trace fidelity of the target gate. In particular, we use the L-BFGS-B optimizer implemented by SciPy~\cite{2020SciPy-NMeth} with the cost function gradient calculated by automatic differentiation enabled by JAX~\cite{jax2018github}.

A schematic is provided in Fig.~\ref{fig:pulse_gen_schematic} to visualize the control of the qudit-qubit system using the parameterized pulse. We show an example of the optimized pulses (solid red and green lines) in the bottom right compared to the standard implementation (dashed lines), and the B-spline pulse parameters representing the optimized pulses in the upper right.

\begin{figure}[h!]
    \centering
    \includegraphics[width=1.0\columnwidth]{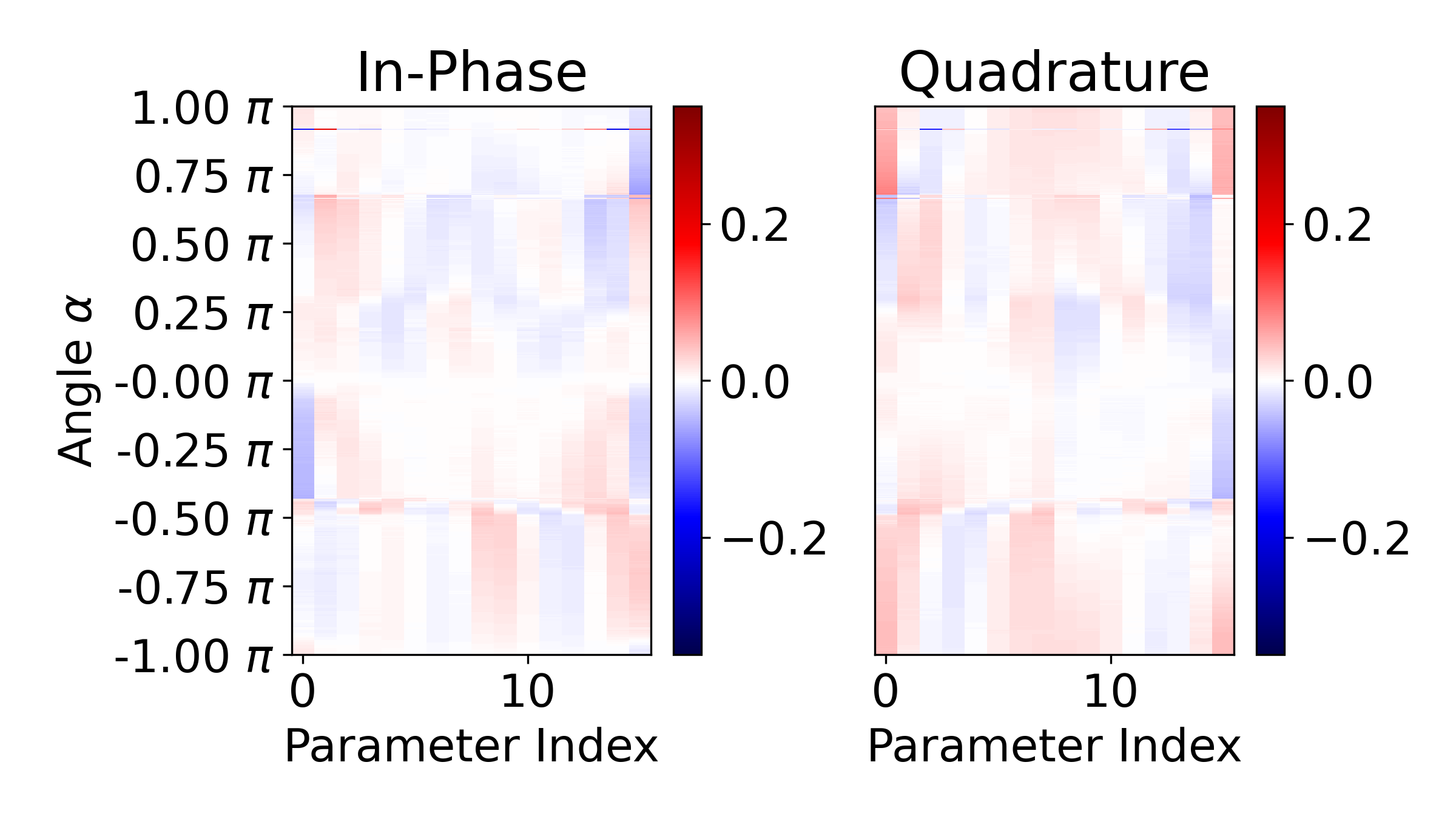}
    \caption{Pulse Parameter Heatmap}
    \label{fig:data-heatmap}
\end{figure}

For training, validation and testing purposes, a data set describing the pulse parameters of 10,000 different input angles $\alpha$ is generated for a SNAP gate with phase shift on the second level ($n=2$) of a qudit with five logical levels ($d=5$). The data set is visualized in Fig.~\ref{fig:data-heatmap}. Although it is beyond the scope of this paper, the same optimal control protocol described above can also be used to generate SNAP gates operating on other levels or in qudits with various logical dimensions.

Each parameter set corresponding to a distinct phase shift angle is optimized with initial parameters motivated by the standard SNAP implementation. There, a geometric phase is selectively applied at the qudit level $n$ by two qubit rotations, and the pulses realizing the two rotations can be determined analytically~\cite{quantum_rf_cavity_control_gate,Heeres_2015}.

Note that the standard SNAP implementation requires the pulse duration time to be much longer than $1/\xi$. The optimal pulse implementation we consider here has a much shorter duration (290 ns) than that required by the standard implementation ($\sim$2000 ns).
This shorter duration is important in terms of reducing decoherence, which can be viewed as an undesired quantum to classical collapse with a characteristic half-life and introduces errors into the computation.
The shorter duration is also useful in terms of effecting a higher overall computational ``clock-rate.''

\subsection{Machine Learning Model Development}
\label{sub:ml-development}

The development of the software-based ML model involves four steps that are outlined in this section: data preprocessing, model architecture selection, model optimization, and design space exploration. The model is trained using a traditional machine learning framework such as TensorFlow and the focus is on creating and fine-tuning it in its software form to achieve optimal performance, specifically minimizing infidelity. This trained model serves as a reference and specification for the hardware implementation. 

\subsubsection{Data Pre-processing}
\label{sec:preprocessing}
The data produced by our numerical simulation described in Sec.~\ref{sec:data} is shown in Fig.~\ref{fig:data-heatmap}. The 16 parameters $\theta$ for the in-phase pulse and the quadrature pulse are on the $x$ axis, and the full range of phase angles $\alpha \in (-\pi, \pi)$ is on the $y$ axis.
A heat map represents the parameter values. After inspecting the figure carefully, one can make the following observations.

\begin{itemize}\setlength{\itemsep}{-0.15em}
    \item A small number of sample parameters have a significantly larger magnitude than the remainder. Furthermore, discontinuities appear as bands across parameters at some select angle values.
    \item Along the angle axis, regions are present where fast switching between negative and positive parameter values occurs.
\end{itemize}

Both of these phenomena introduce discontinuities, which make it harder to achieve good ML performance. As such, we propose the following preprocessing steps:

\begin{itemize}\setlength{\itemsep}{-0.15em}
    \item Filter out the parameter values for which the classical optimization procedure results in infidelity exceeding a certain threshold. Empirically, $10^{-4}$ was found to be a good choice.
    \item Perform a smoothing operation along the angle axis, using a uniform windowed average over 50 samples (equivalent to 1.8$^{\circ}$). 
\end{itemize}

\begin{figure}[t]
    \centering
    \includegraphics[width=1.0\columnwidth]{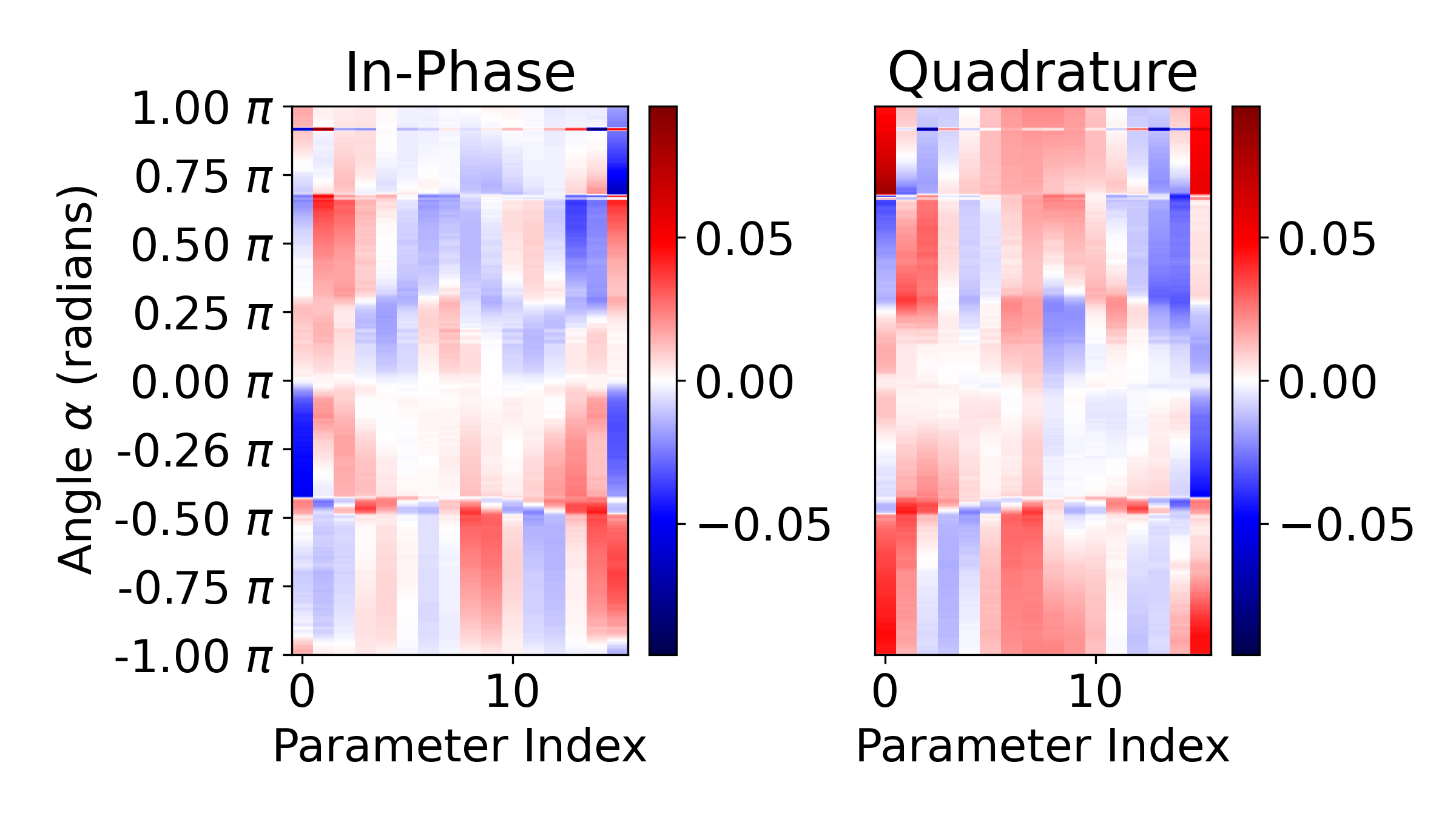}
    \caption{Pre-processed Pulse Parameter Heatmap}
    \label{fig:preprocessed-data-heatmap}
\end{figure}

\begin{figure}[t]
    \centering
    \includegraphics[width=1.0\columnwidth]{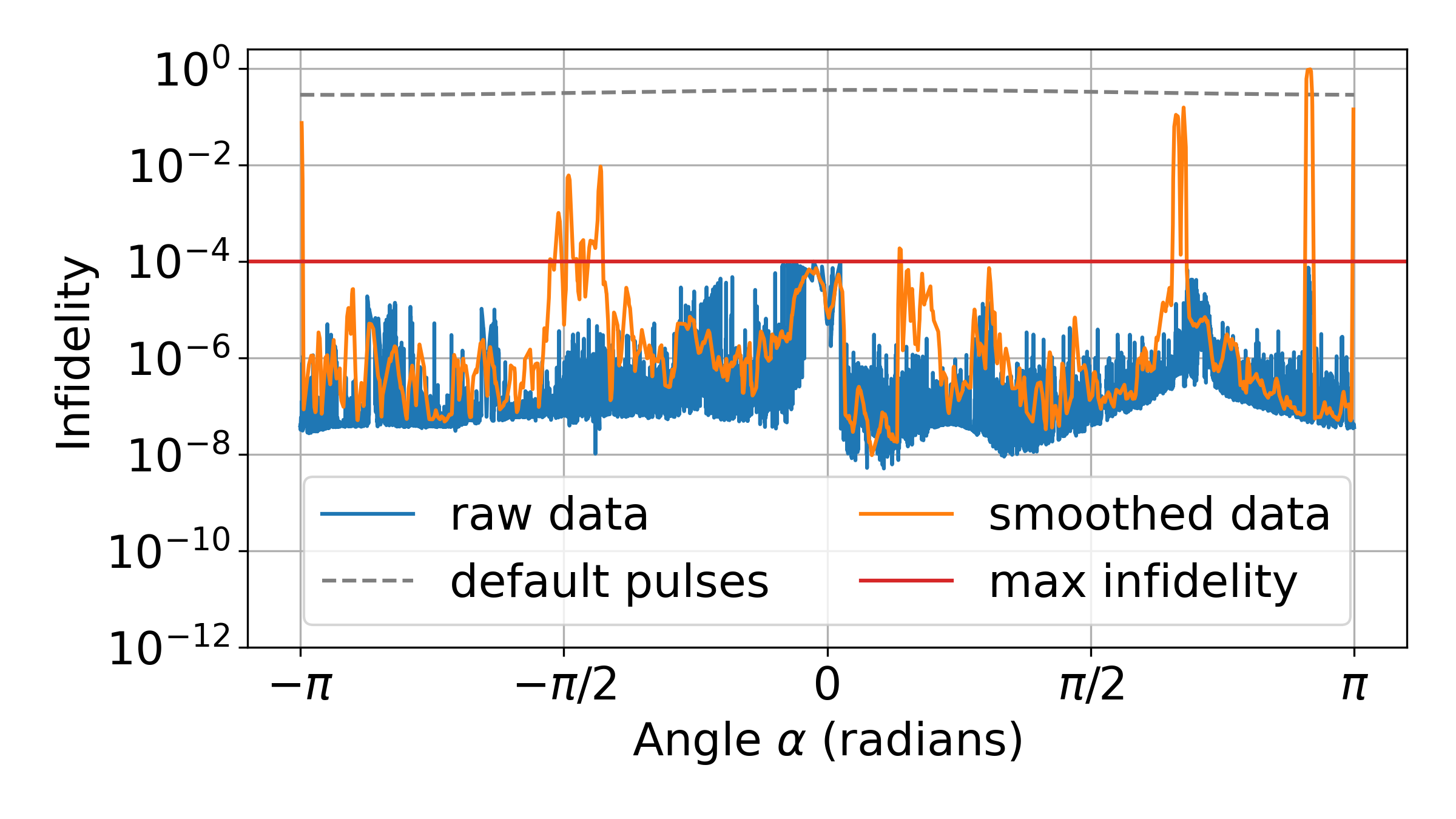}
    \caption{Training Data Infidelity}
    \label{fig:data-infidelity}
\end{figure}

The pulse parameters after pre-processing are shown in Fig.~\ref{fig:preprocessed-data-heatmap}.
The reduced variation in the parameter values is confirmed by a narrower range of values as compared to Fig.~\ref{fig:data-heatmap}, which featured larger outliers.

The infidelity performance metric of the training data, both before and after smoothing, is shown in Fig.~\ref{fig:data-infidelity}.
The smoothing operation does have an adverse effect on the infidelity in regions where rapid changes occur (e.g., around $\alpha \simeq -0.5 \pi$ and $\alpha \simeq 0.75 \pi$).
The gray dashed line in the figure represents the infidelity of the default pulse parameters used as the initial guesses of the optimal control described in Sec.~\ref{sec:data}. Initial guesses are motivated by the standard SNAP implementation, and the high infidelity indicates that optimal control pulses are necessary for a short gate time.

Finally, we split the data set into training, validation, and testing sets, respectively, which contain 80\%, 10\% and 10\% of the data samples.

\subsubsection{Model Architectures}
\label{subsub:architectures}

For this work, we developed three main machine learning model architectures: a multilayer perceptron (MLP), two variations of the Mixture-of-Experts (MoE) model, and compact MLPs derived via knowledge distillation. The distilled models retain high fidelity while significantly reducing complexity, making them well-suited for deployment on resource-constrained hardware platforms.

\textbf{\textit{Multi-Layer Perceptron}}.
The first and more elementary architecture is an MLP, which, in essence, implements our target functionality using several matrix multiplication layers, each passed through non-linear activation functions.
According to the universal approximation theorem, such networks can approximate all continuous multivariate functions~\cite{universal_approximation_theorem}. 
Although recently, it has been proven that the universal approximation theorem holds for discontinuous functions as well, as long as the network is at least three layers deep, in practice, MLPs struggle with such cases~\cite{discontinuous_approximation_theorem}.
In our case, Fig.~\ref{fig:preprocessed-data-heatmap} clearly shows discontinuous behavior in some regions (e.g., around $\alpha \simeq -0.5 \pi$ and $\alpha \simeq 0.75 \pi$).

\textbf{\textit{Mixture-of-Experts}}.
\begin{figure}[t]
    \centering
    \includegraphics[width=0.45\columnwidth]{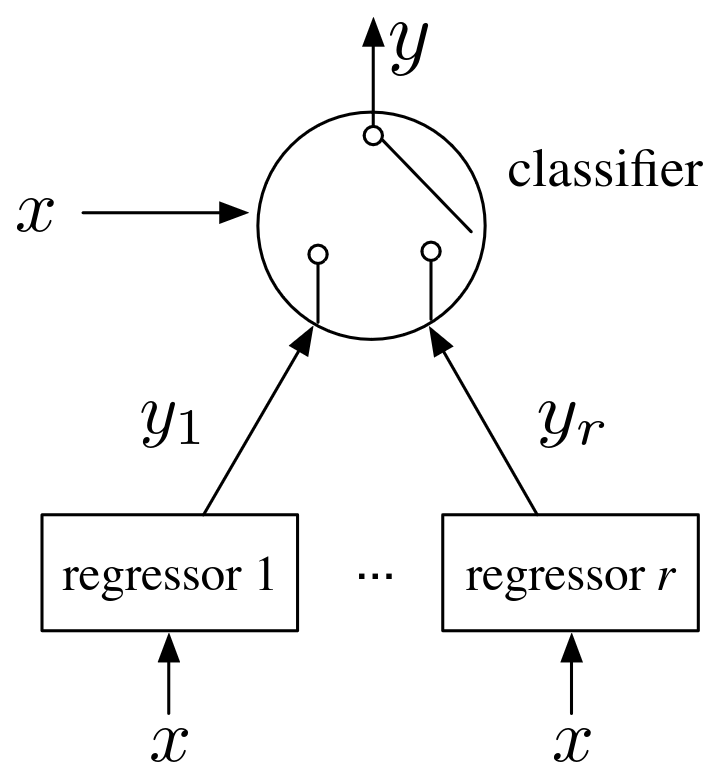}
    \caption{Mixture-of-Experts Model Architecture, from \cite{moe_nonlinear_control}}
    \label{fig:moe-architecture}
\end{figure}
The authors of \cite{moe_nonlinear_control} tackle the issue of discontinuous non-linear optimal control with machine learning models in the context of drone obstacle avoidance. Specifically, they propose using the MoE architecture to approximate functions, as shown in Fig.~\ref{fig:moe-architecture}. The core idea is that each expert can learn the outputs of a continuous function subdomain and that a selector or classifier can then combine their outputs.

In our work, we implement two versions of the MoE architecture.
\begin{itemize}
    \item An ensemble of MLP regressors with a MLP with softmax activation function as a classifier, further labeled \emph{MoE}. We refer to this approach as ``soft switching.''
    \item An ensemble of MLP regressors with a hard (0 or 1) model selector and hand-defined regions as a classifier, further labeled \emph{Multi-Region (MR)}. Owing to the hand-selected switching, we refer to this as ``hard switching.'' Note the contrast to soft switching where a weighted average is implicitly learned.
\end{itemize}

The advantage of these architectures is that the network can explicitly model or learn the discontinuities in the data. However, this comes at the cost of the need for multiple regressor instances and, thus, an increased total model size. The next section will describe our hyperparameters and techniques for model optimization during training, while Sec.~\ref{subsub:model-dse} further explores the trade-off between the three model architectures.
Soft switching allows the NN to learn which parameters are best for a given set of angles, but hard switching may be more efficient in terms of resource use on hardware.

\subsubsection{Model Optimization}
\label{subsub:optimization}

To train our models, we employ the following two-step approach.
The first step is relatively fast compared to the more accurate second step, and serves as a quick way to get a solution in the right area of parameter space.
First, we perform traditional ML training on the data set described in Sec.~\ref{sec:data}.
We use TensorFlow and the Adam optimizer with mean squared error (MSE) as a loss function, comparing predicted pulse parameters to values produced by the ML algorithm.
When training, we find it effective to lower the learning rate as the training progresses, so we adopt an exponentially decaying learning rate schedule that evolves from $10^{-3}$ to $10^{-5}$~\cite{li2019exponential}. We train each model for 50 epochs and retain the version that achieves the lowest validation loss throughout the training process.

For the second step, we use the special property of the infidelity cost function, which we used in our initial data generation: it is itself differentiable.
We can exploit this property to optimize the model directly towards this measure rather than the MSE.
Although significantly more expensive to training using the infidelity directly because of the required simulation costs, it may be used as a fine-tuning step with a limited number of randomly generated angles.
Furthermore, we expect this method to be more appropriate for working with real hardware in cases where the ``simulation'' becomes nature, and the response to a given set of pulse parameters is no longer precisely described by the simulation.
In order to maximize learning, the probability to select a certain angle is initially chosen to be proportional to the model infidelity, thus prioritizing angles with poor performance.
As time goes on, this probability distribution is flattened towards uniform selection.
Using the Adam optimizer with a fixed learning rate of $10^{-3}$, fine-tuning is performed on the trained model for 25 batches of each 16 angles per round of fine-tuning.

\subsubsection{Design Space Exploration}
\label{subsub:model-dse}

When translating ML models from software to hardware implementations on programmable logic devices such as eFPGAs, careful model tuning becomes crucial to navigate the trade-offs between latency, resource utilization, and overall performance. This involves optimizing the design to fit within the limited hardware resources while maintaining acceptable levels of accuracy and throughput, which often requires quantization, pruning, and parallelization techniques tailored to the target architecture.
These considerations prompted our exploration of ML model architectures and training hyperparameters to achieve Pareto-optimal latency and resource utilization.

\textbf{\textit{MLP Design Space Exploration}}.
To comprehensively assess MLP modeling capabilities, we randomly generated multiple model configurations featuring different depths and numbers of neurons per hidden layer. 
%
We generated a total of 100 configurations. We trained each model using the hyper-parameters and details described as the first step of Sec.~\ref{subsub:optimization}.
Initially, we did not apply fine-tuning; instead, we just used MSE loss performance to identify Pareto configurations with the best modeling ability.
Then, we selected 11 configurations for which the increase in parameter count yielded the biggest improvement in model performance. We ran the fine-tuning procedure for ten iterations to further improve the model performance.
These models, their test loss before fine-tuning, and their mean and maximum infidelity performance measures after fine-tuning are listed in Table~\ref{tab:model-ft-selection}. We named each MLP model at this stage with its total parameter count.

\begin{table}
\centering
\resizebox{\columnwidth}{!}{
\begin{tabular}{ccccc}
\toprule
\textsc{Model} & \textsc{Num. of} & \textsc{Test}  & \textsc{Mean}       & \textsc{Max} \\
\textsc{Name}  & \textsc{Params}  & \textsc{Loss}  & \textsc{Infidelity} & \textsc{Infidelity} \\
\midrule
\texttt{mlp\_100}     & 100             & 0.000324           & 0.056265          & 0.267205         \\
\texttt{mlp\_106}    & 106             & 0.000181           & 0.230588          & 0.500094         \\
\texttt{mlp\_112}    & 112             & 0.000181           & 0.232101          & 0.494881         \\
\texttt{mlp\_182}    & 182             & 0.000180           & 0.051570          & 0.212557         \\
\texttt{mlp\_194}    & 194             & 0.000124           & 0.051655          & 0.216860         \\
\texttt{mlp\_494}    & 494             & 0.000113           & 0.033348          & 0.195251         \\
\rowcolor{gray!20}
\texttt{mlp\_514}    & 514             & 0.000101           & 0.014049          & 0.066427         \\
\texttt{mlp\_1022}    & 1022            & 0.000100           & 0.035468          & 0.190022         \\
\texttt{mlp\_1320}     & 1320            & 0.000060           & 0.009029          & 0.037378         \\
\texttt{mlp\_1608}     & 1608            & 0.000041           & 0.005823          & 0.026224         \\
\rowcolor{gray!20}
\texttt{mlp\_4308}    & 4308            & 0.000034           & 0.005083          & 0.010809         \\ 
\bottomrule
\end{tabular}
}
\vspace{10px}
\caption{Eleven MLP models out of an initial 100 models we selected for design space exploration. In general, mean squared error (MSE) and infidelity decrease as the number of model parameters increases. Highlighted the best performing model and the best performing model under 1,000 parameters that we chose for further exploration.}
\label{tab:model-ft-selection}
\end{table}

\begin{figure}
    \centering
    \includegraphics[width=0.85\columnwidth]{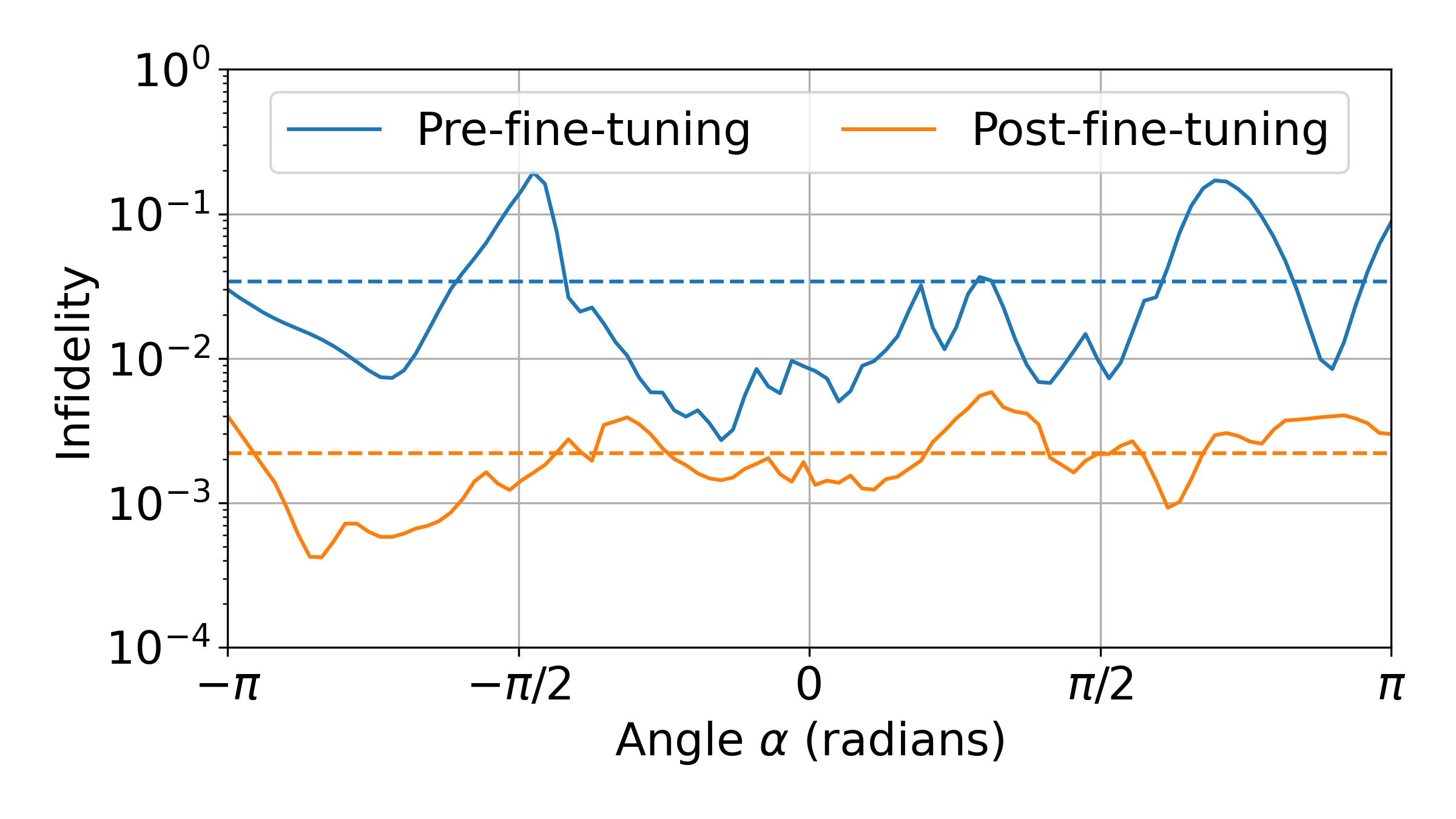}
    \caption{Model \texttt{mlp\_4308} Infidelity Performance}
    \label{fig:random30-infidelity-comparison}
\end{figure}

\begin{figure}
    \centering
    \includegraphics[width=0.85\columnwidth]{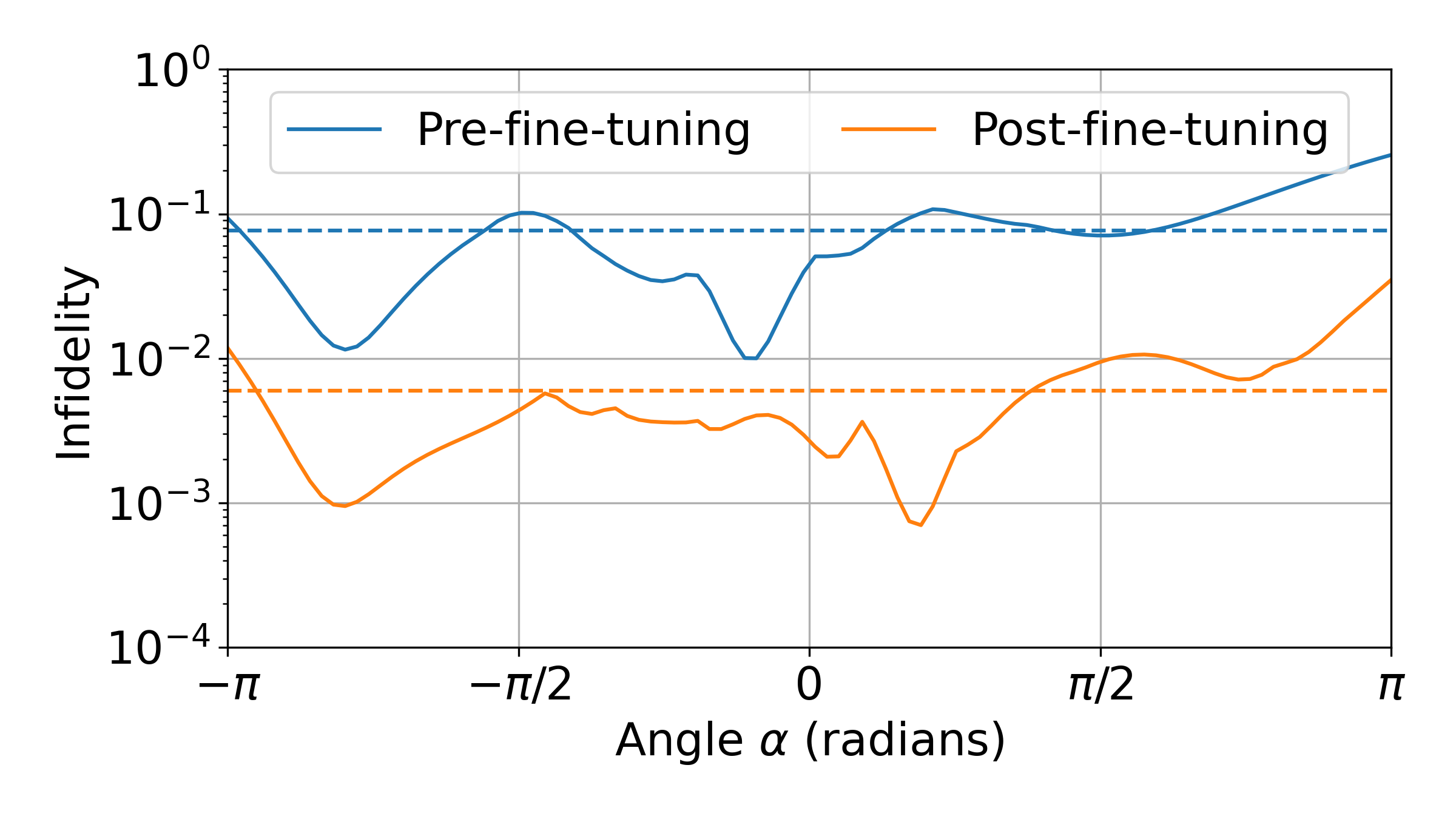}
    \caption{Model \texttt{mlp\_514} Infidelity Performance}
    \label{fig:random67-infidelity-comparison}
\end{figure}

Finally, we investigate how infidelity could be further reduced by using longer fine-tuning sessions.
We perform fine-tuning for 20 more iterations on the models \texttt{mlp\_4308}, the overall best performing MLP, and \texttt{mlp\_514}, the best performing MLP with less than 1,000 parameters.
Fig.~\ref{fig:random30-infidelity-comparison} and Fig.~\ref{fig:random67-infidelity-comparison} provide performance comparisons for these models before (blue) and after fine-tuning (orange).
These examples show that the proposed fine-tuning procedure effectively improves the infidelity of the trained models.


\begin{figure}[b]
    \centering
    \includegraphics[width=0.85\columnwidth]{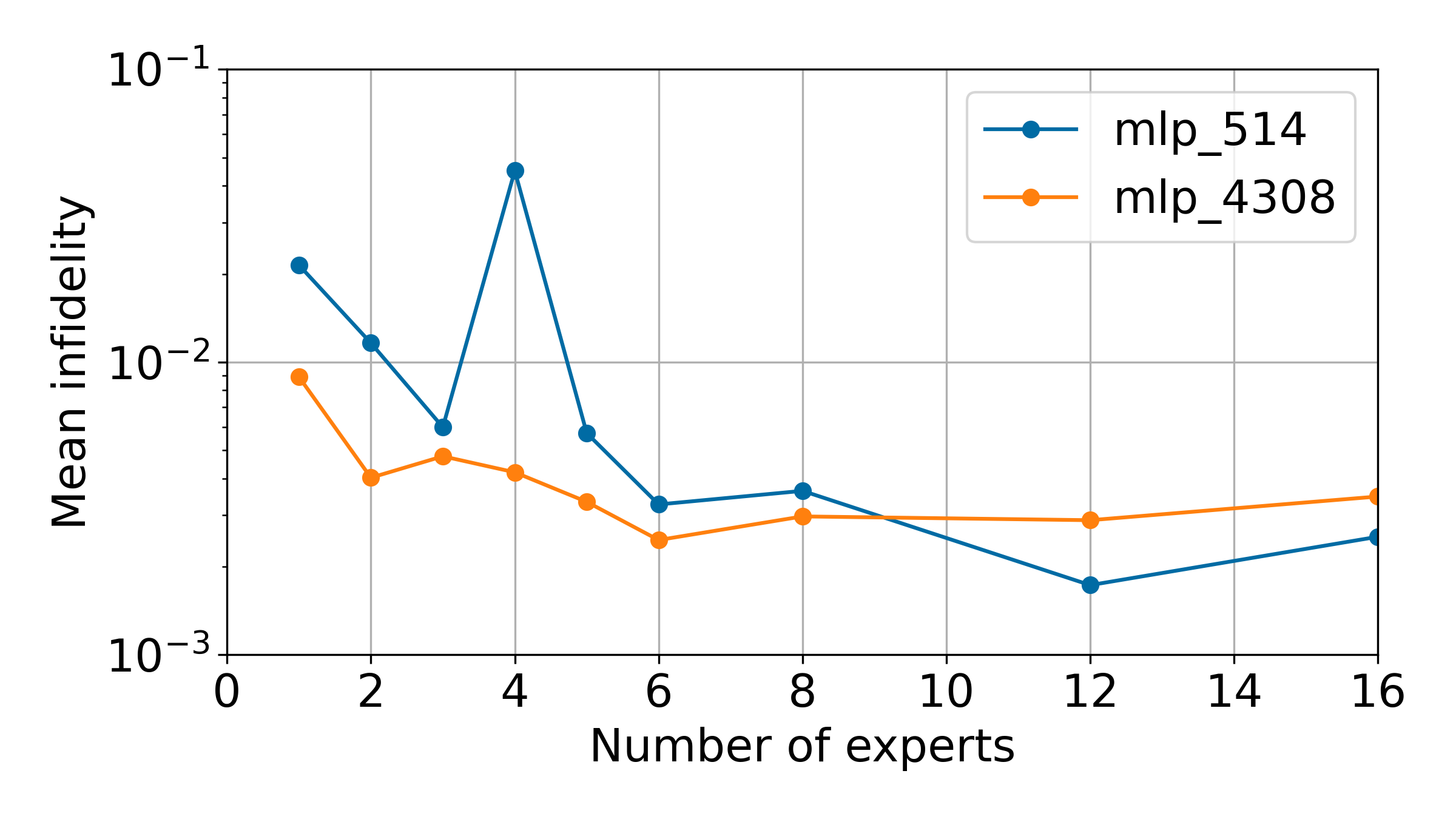}
    \caption{MoE expert count vs. model infidelity}
    \label{fig:nr-of-experts-influence}
\end{figure}

\textbf{\textit{Mixture-of-Experts Design Space Exploration}}
Next, we assess the potential of using a MoE architecture to model the qudit control application.
We implement MoE models with the previously selected \texttt{mlp\_514} and \texttt{mlp\_4308} configurations as regressors and vary the number of experts from 1 to 16.
Each configuration undergoes the standard training procedure and five iterations of fine-tuning. 
The results, visualized in Fig.~\ref{fig:nr-of-experts-influence}, show a general improvement in performance as the number of experts increases. We also observe that with more than six experts, the reduction in infidelity per added expert is minimal.

The authors of \cite{moe_nonlinear_control} show that a well-selected set of regions in the data, for which the outputs are each predicted by a separate regressor, can outperform the softmax output combination in the traditional MoE architecture. This is the core idea behind the architecture we call Multi-Region (MR). By analyzing the mean first derivative of the pulse parameters over the range of angles, we choose five approximately continuous subdomains separated by the following phase angles $\alpha$: [$-0.490 \pi$, $-0.426 \pi$, $0$, $0.682 \pi$]. 

Consequently, we built two MR models with these five decision regions. As regressors, we pick the \texttt{mlp\_514} and \texttt{mlp\_4308} MLP models. Both models are trained as explained in Sec.~\ref{subsub:optimization} and fine-tuned for 30 iterations. For comparison, we used MoE models with five experts in Fig.~\ref{fig:nr-of-experts-influence} and fine-tuned them equally to 30 iterations.

\begin{figure}[t]
    \centering
    \includegraphics[width=0.85\columnwidth]{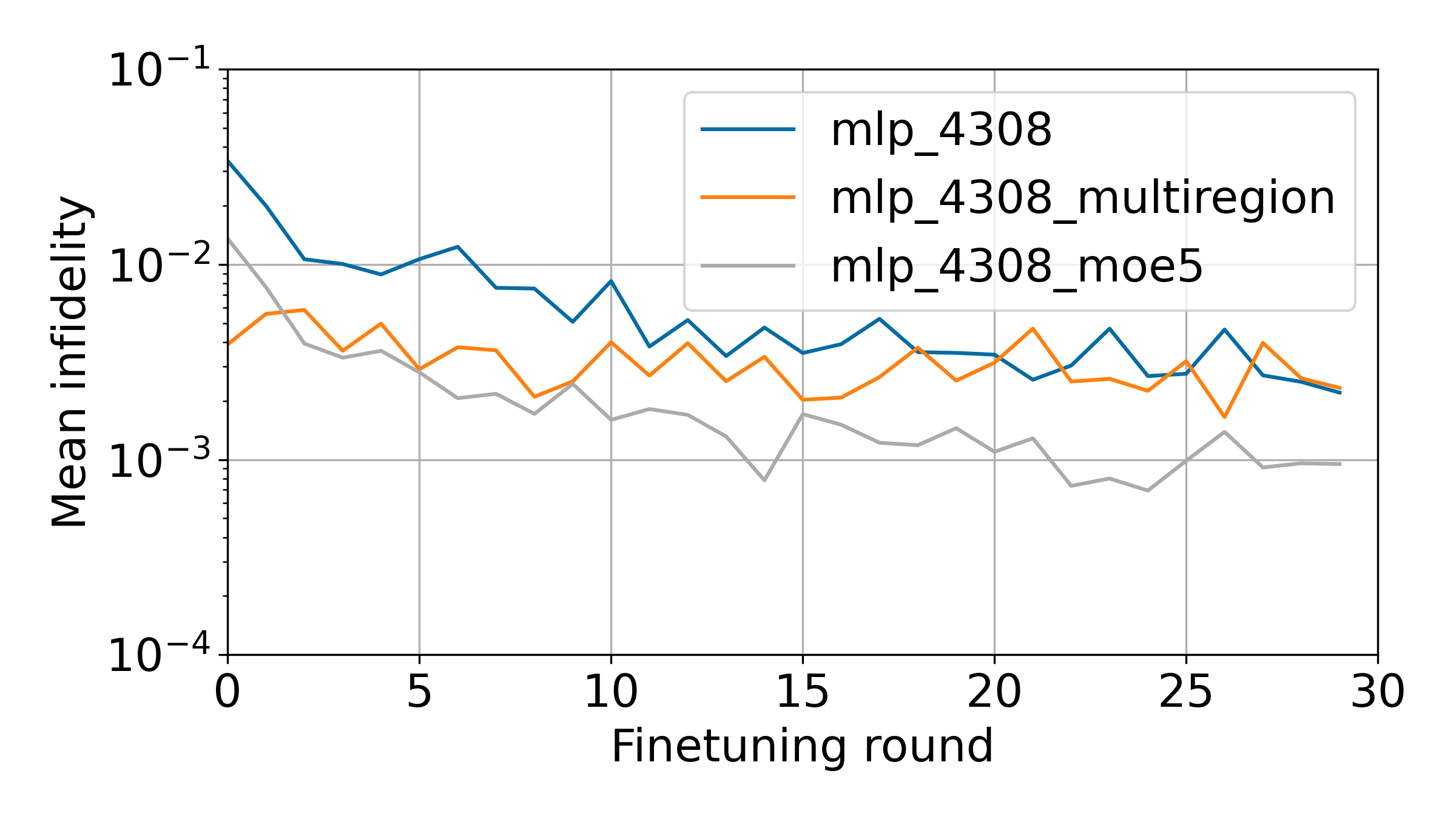}
    \caption{MoE Fine-tuning for \texttt{mlp\_4308}}
    \label{fig:multiregion-random30-results}
\end{figure}

\begin{figure}[t]
    \centering
    \includegraphics[width=0.85\columnwidth]{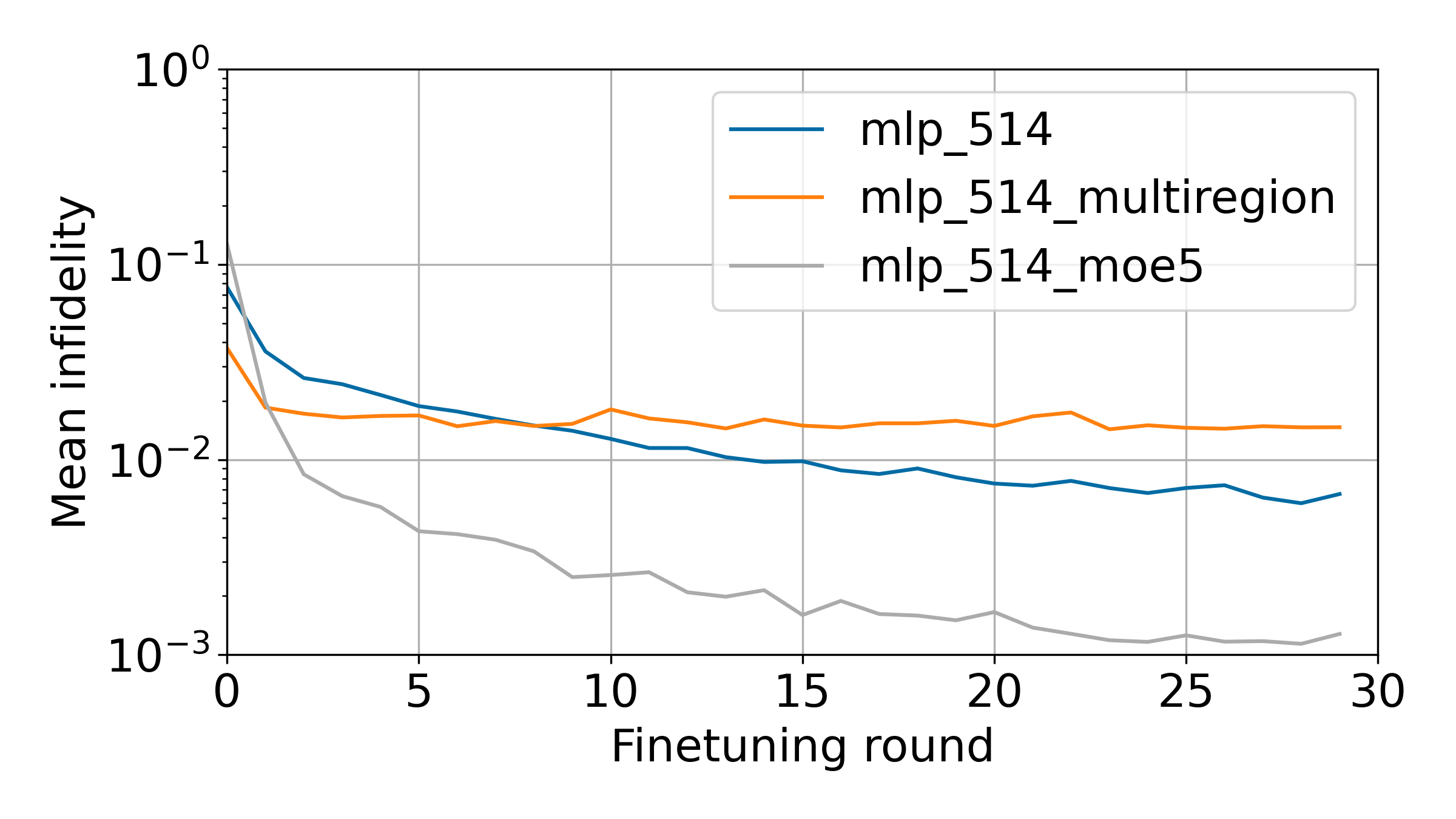}
    \caption{MoE Fine-tuning for \texttt{mlp\_514}}
    \label{fig:multiregion-random67-results}
\end{figure}

The results are provided in Fig.~\ref{fig:multiregion-random30-results} and Fig.~\ref{fig:multiregion-random67-results} for the \texttt{mlp\_4308} and \texttt{mlp\_514} architectures, respectively.
MoE architectures outperform the base models by a respectable margin, yielding a mean infidelity below $10^{-3}$ for the \texttt{mlp\_4308} configuration. However, the fine-tuning process for the MR models yields only limited improvements. Further investigation shows this to be the consequence of the hard-switching action between the different regressors, which results in unstable learning for smaller batch sizes. As a workaround, we evaluated the results of every intermediate fine-tuning step and manually composed a model with the best regressor found for each decision region.

\begin{figure}[t]
    \centering
    \includegraphics[width=0.85\columnwidth]{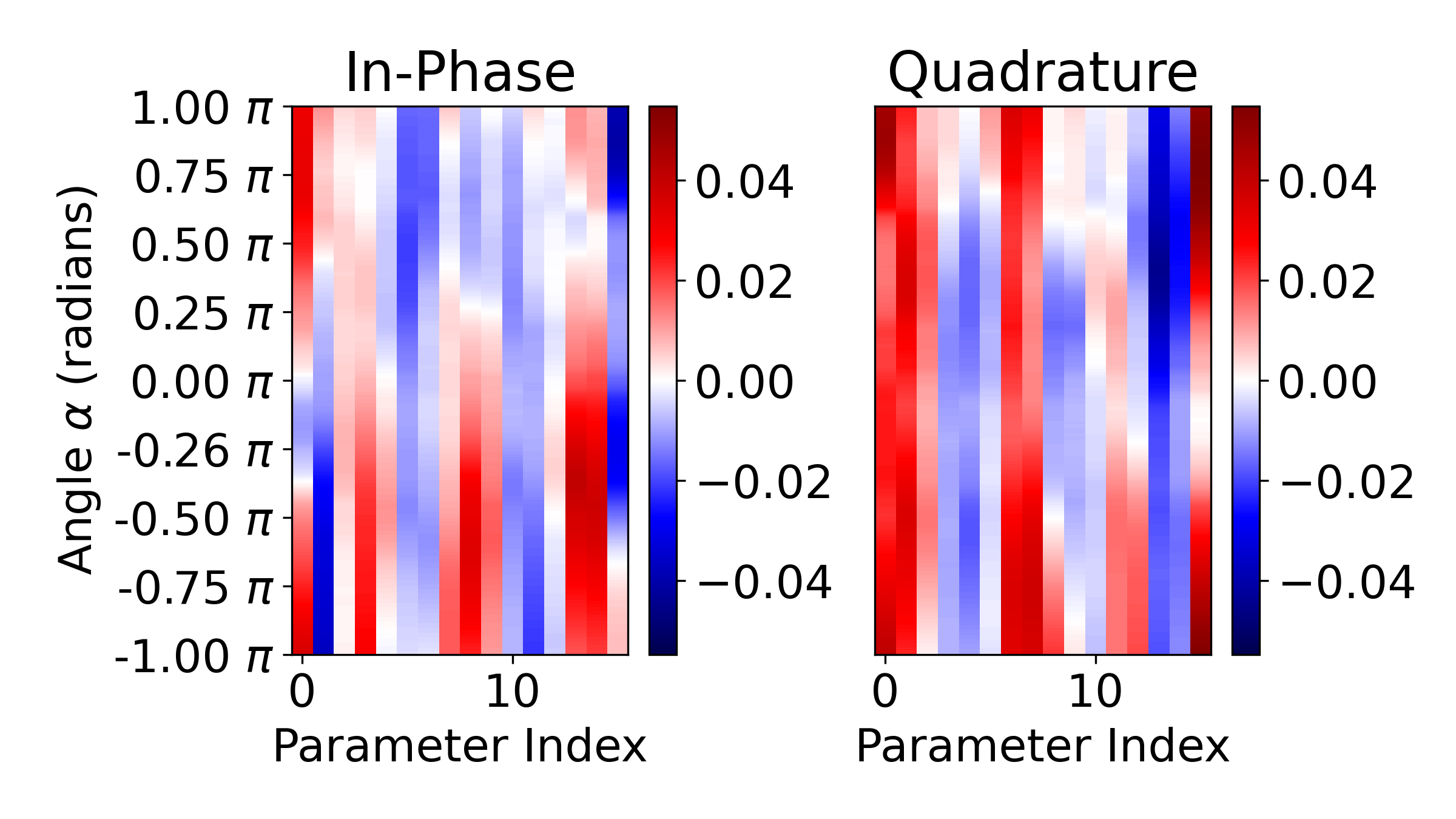}
    \caption{MoE \texttt{mlp\_4308} Pulse Parameter Heatmap}
    \label{fig:random-30-moe5-heatmap}
\end{figure}

\begin{figure}[t]
    \centering
    \includegraphics[width=0.85\columnwidth]{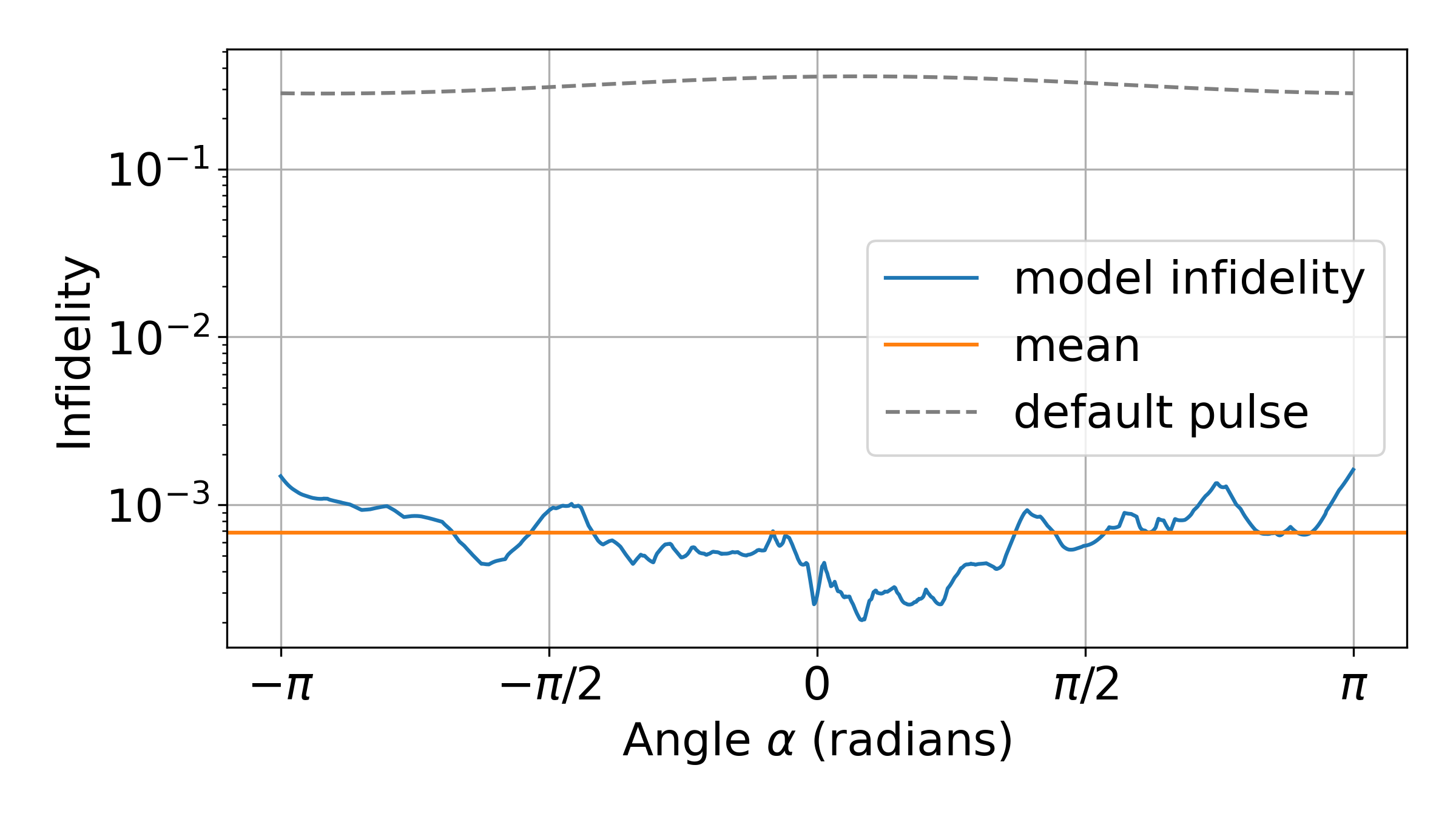}
    \caption{MoE \texttt{mlp\_4308} Infidelity Performance}
    \label{fig:random-30-moe5-infidelity}
\end{figure}

\textbf{\textit{Knowledge Distillation}}.
Fig.~\ref{fig:random-30-moe5-heatmap} shows the pulse parameter heatmap of our best performing model so far (MoE based on \texttt{mlp\_4308} with five experts).
Compared with Fig.~\ref{fig:preprocessed-data-heatmap}, it is very interesting to note that the pulse parameters found by this model look very different from the original data set. Rather, throughout the fine-tuning process, the model produces a set of continuous parameters over all input angles. At the same time, the model balances infidelity across all angles, resulting in a mean value below $10^{-3}$, as shown in Fig.~\ref{fig:random-30-moe5-infidelity}.

Although these are promising results, there are two issues with this model.
First, because the model and the classifier network need multiple regressors, the total number of parameters of the MoE model is more than five times that of its base regressor. Furthermore, complex architectures such as MoE models are more difficult to implement in hardware than standard MLPs. 

The continuous character of the predicted parameters allows for the use of knowledge distillation (KD), a technique in which the knowledge of a MoE is learned by a single model \cite{knowledge_distillation}.
This method takes a two step approach:

\begin{enumerate}
    \item Create a dataset containing input-output samples produced by the MoE model. 
    In this case, we go for 10,000 linearly spaced input angles.
    \item Train a MLP model on this new dataset. 
    Here we select all configurations from Table \ref{tab:model-ft-selection}.
\end{enumerate}

After training, KD models with parameter counts exceeding 1000 surpass their fine-tuned counterparts in mean infidelity performance, confirming the effectiveness of the KD technique. However, the results for models with lower parameter counts do not achieve the same performance as their fine-tuned counterparts. The reason for this is likely that they lack sufficient modeling capacity to capture the trends in the KD data set.

\begin{figure}[t]
    \centering
    \includegraphics[width=1.0\columnwidth]{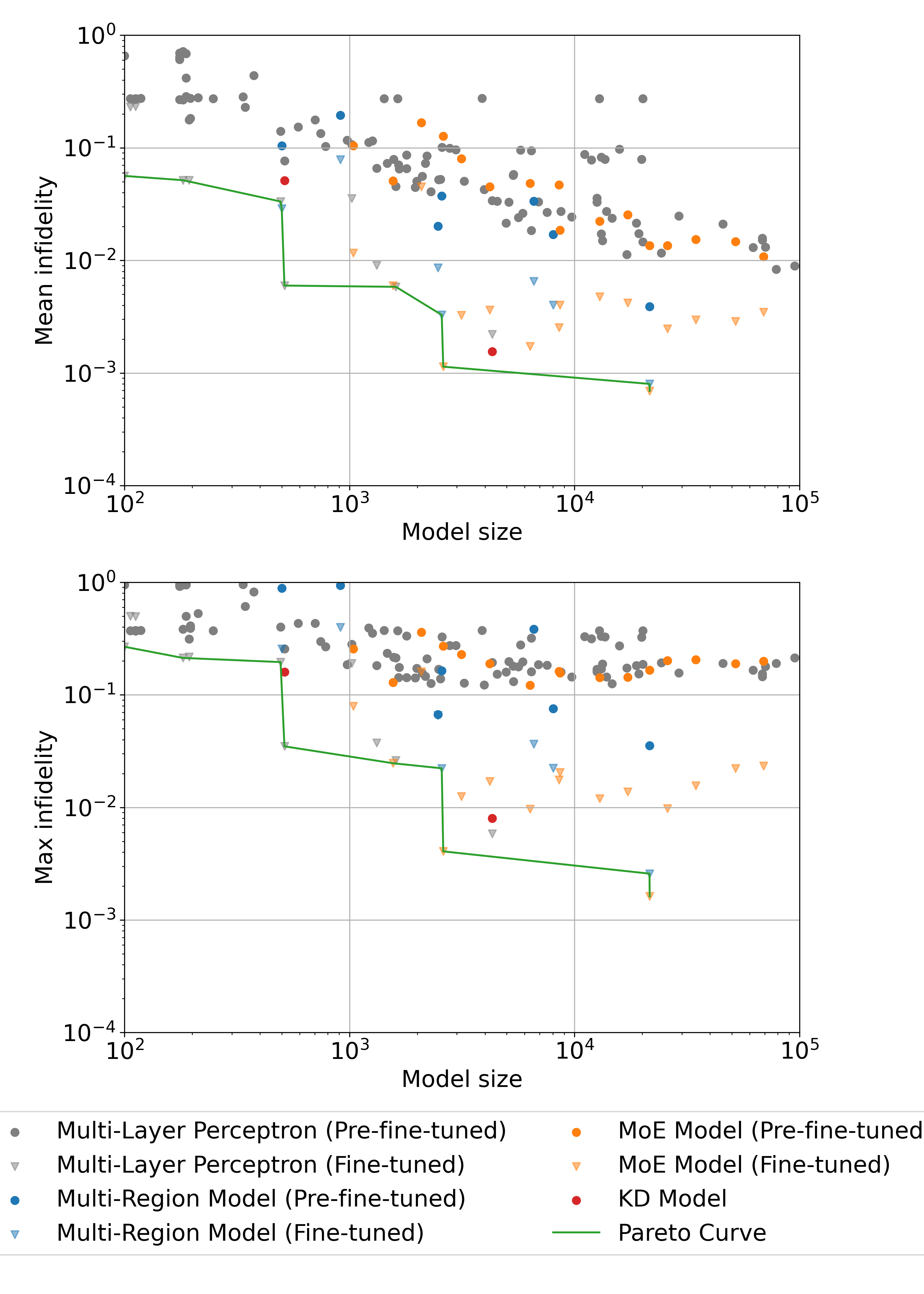}
    \caption{Design Space Exploration Overview}
    \label{fig:dse-overview}
\end{figure}

\textbf{\textit{DSE Results}}.
Fig.\ref{fig:dse-overview} presents an overview of all configurations for each model architecture explored during the design space exploration process for both mean and maximum infidelity performance measures. We show the Pareto curve for each. Furthermore, a distinction is made between results before and after model fine-tuning.

The influence of fine-tuning the models directly on the infidelity cost function is clear, as improvements of more than an order of magnitude can be observed in many cases. MoE models occupy a significant portion of the Pareto curve, especially for larger parameter counts. The MLP architectures often come close to Pareto performance after fine-tuning or through knowledge distillation. For example, \texttt{mlp\_1608}, which is the MLP model that lies closest to the Pareto curve after knowledge distillation and \texttt{mlp\_4308}, which is the best performing MLP model.

Table \ref{tab:pareto-models} provides an overview of the model architectures on the Pareto curve for mean and maximum infidelity and achieves a mean infidelity lower than $10^{-2}$.
The table gives their name, architectural type, parameters, number of fine-tuning rounds (FT), number of parameters, the best achieved mean infidelity, and the corresponding maximum infidelity.
Both MR models on the Pareto curve are those assembled from different fine-tuning checkpoints, indicated by the *-character in the table.

\begin{table}
\centering
\resizebox{\columnwidth}{!}{
\begin{tabular}{|l|c|c|c|c|c|}
\hline
\textbf{Name}           & \textbf{Type} & \textbf{FT} & \textbf{Params} & \textbf{Mean Inf} & \textbf{Max Inf} \\ \hline
\texttt{mlp\_514}        & MLP           & 29          & 514             & 0.005974          & 0.034849         \\ \hline
\texttt{mlp\_514}\_multiregion & MR            & 30*         & 2570            & 0.003278          & 0.022238         \\ \hline
\texttt{mlp\_514}\_moe5        & MoE (5)           & 29          & 2610            & 0.001137          & 0.004068         \\ \hline
\texttt{mlp\_4308}\_multiregion & MR            & 30*         & 21540           & 0.000800          & 0.002580         \\ \hline
\texttt{mlp\_4308}\_moe5        & MoE (5)       & 25          & 21580           & 0.000694          & 0.001621         \\ \hline
\end{tabular}
}
\vspace{10px}
\caption{Pareto Models}
\label{tab:pareto-models}
\end{table}

\subsection{Quantization} 
\label{sec:quantization}

To translate a machine learning model for implementation in hardware, the model first needs to be quantized. 
This section discusses our approach of using QKeras for initial quantization and HLS4ML for making fine adjustments to the layer precisions.

\subsubsection{Model Quantization}
\label{subsub: quantization-and-pruning}
Quantizing a model involves representing weights, biases, and activation functions using fixed-point numbers.
Generally, two approaches are employed: post-training quantization and pre-training quantization.
In pre-training quantization, the model is developed and trained using fixed-point integers for these parameters. 
Conversely, in post-training quantization, the model's weights are quantized after the training process is completed.
In this report, our focus is on developing a pre-training quantized model, as it allows the model to adapt to lower precision during training, typically resulting in increased accuracy.

\begin{figure}[ht]
    \centering
    \includegraphics[width=0.85\columnwidth]{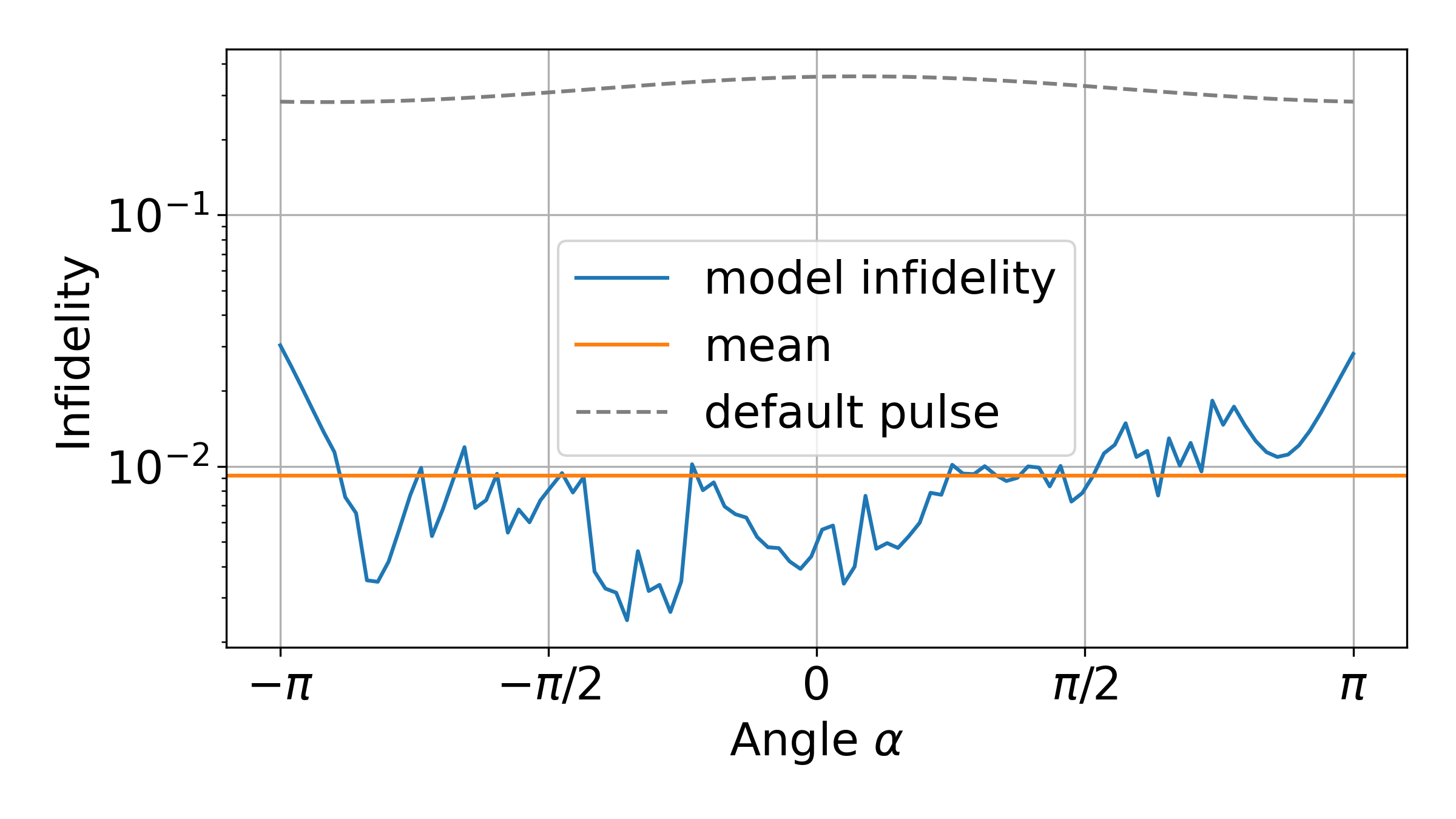}
    \caption{Infidelity vs. Angle for 4-bit precision parameters}
    \label{fig:quantization-infidelity-4}
\end{figure}

\begin{figure}[h]
    \centering
    \includegraphics[width=0.85\columnwidth]{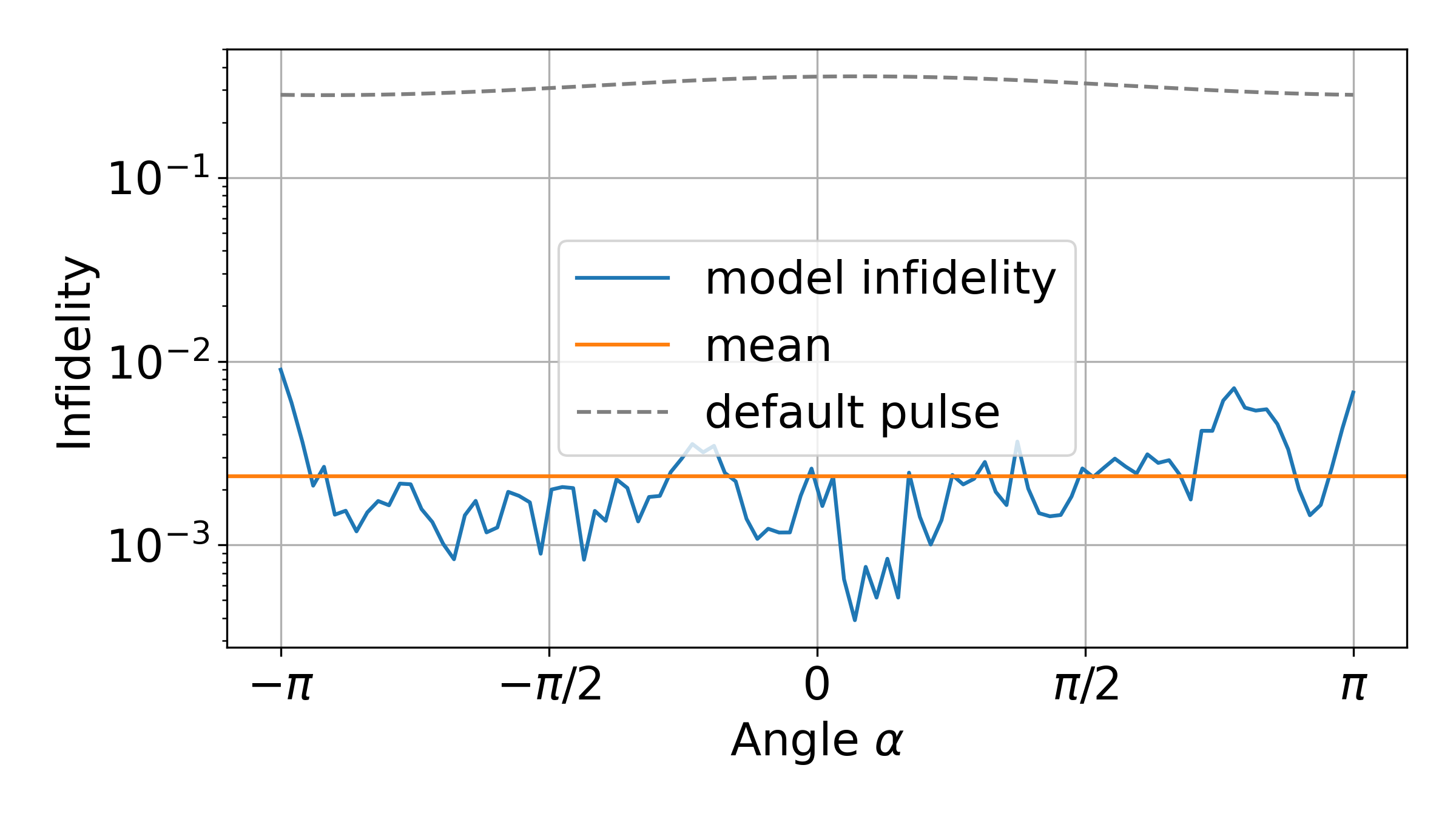}
    \caption{Infidelity vs. Angle for 6-bit precision parameters}
    \label{fig:quantization-infidelity-6}
\end{figure}

\begin{figure}[h]
    \centering
    \includegraphics[width=0.85\columnwidth]{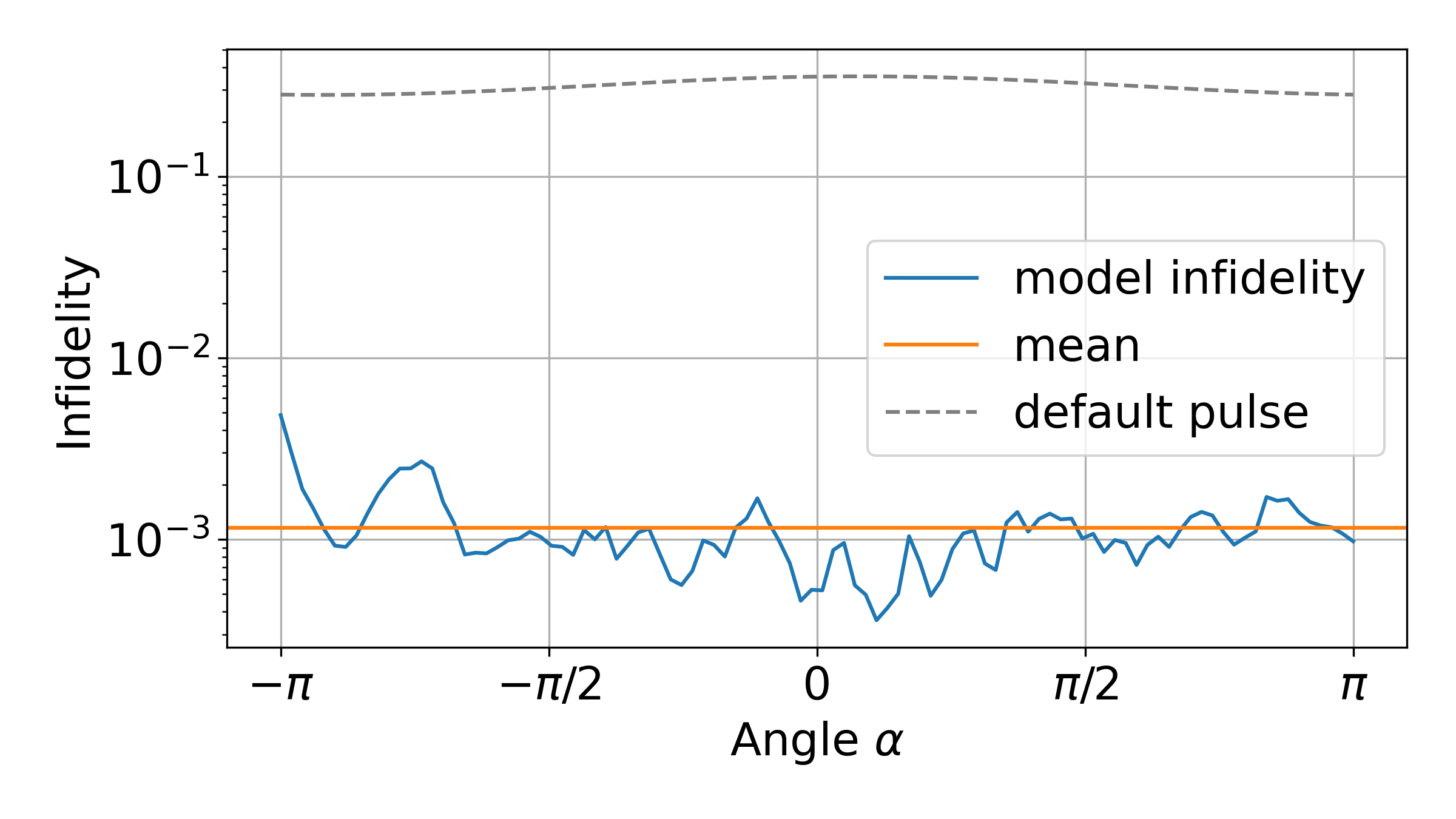}
    \caption{Infidelity vs. Angle for 8-bit precision parameters}
    \label{fig:quantization-infidelity-8}
\end{figure}

We select the random\_3 model with knowledge distillation for quantization and synthesis using HLS. This choice is informed by its relatively low parameter count and its proximity to the Pareto curve in Fig.~\ref{fig:dse-overview}.

Our design exploration begins by quantizing the weights, biases, and ReLU activation functions using QKeras.
This enables us to maintain the same architecture as random\_3 while adjusting the quantization bits for the parameters.
We set the integer bits of weights, biases, and activation functions to zero, as the typical range for trainable parameters in such a multi-layer perceptron (MLP) lies between 1 and -1.
The fractional length of the number of bits varied from 1 to 8.
For bit-sizes 1, 2, and 3, the mean infidelities exceeded $10^{-2}$.
Conversely, for bit-size quantizations 4, 5, 6, 7, and 8, the mean infidelities are below $10^{-2}$. Increasing the bit precision from 4 to 8 bits results in a progressive decrease in infidelity, as illustrated in Figures \ref{fig:quantization-infidelity-4}, \ref{fig:quantization-infidelity-6} and \ref{fig:quantization-infidelity-8}.
Consequently, we decided to proceed further in the hardware implementation process only for parameter bit quantizations of 4 to 8.

\subsubsection{Layer Precision Adjustments}

HLS4ML serves as our high-level synthesis tool which leverages Vivado HLS to convert the QKeras model into Verilog code. 
Providing users with a set of adjustable parameters, HLS4ML empowers fine-tuning of model quantization.
One key setting is the ``results'' parameter, which determines the fixed-point precision of the output of each layer of the neural network. By default, this is set to $⟨16,6⟩$, meaning 16 total bits with 6 fractional bits. Adjusting this precision directly affects hardware resource utilization, numerical accuracy, and timing performance during synthesis. Our experimentation reveals that such high precision is indeed not required; instead, we can achieve comparable results with significantly smaller precision, thereby reducing the utilization of FPGA resources during synthesis.

\begin{figure}[h]
    \centering
    \includegraphics[width=1\columnwidth]{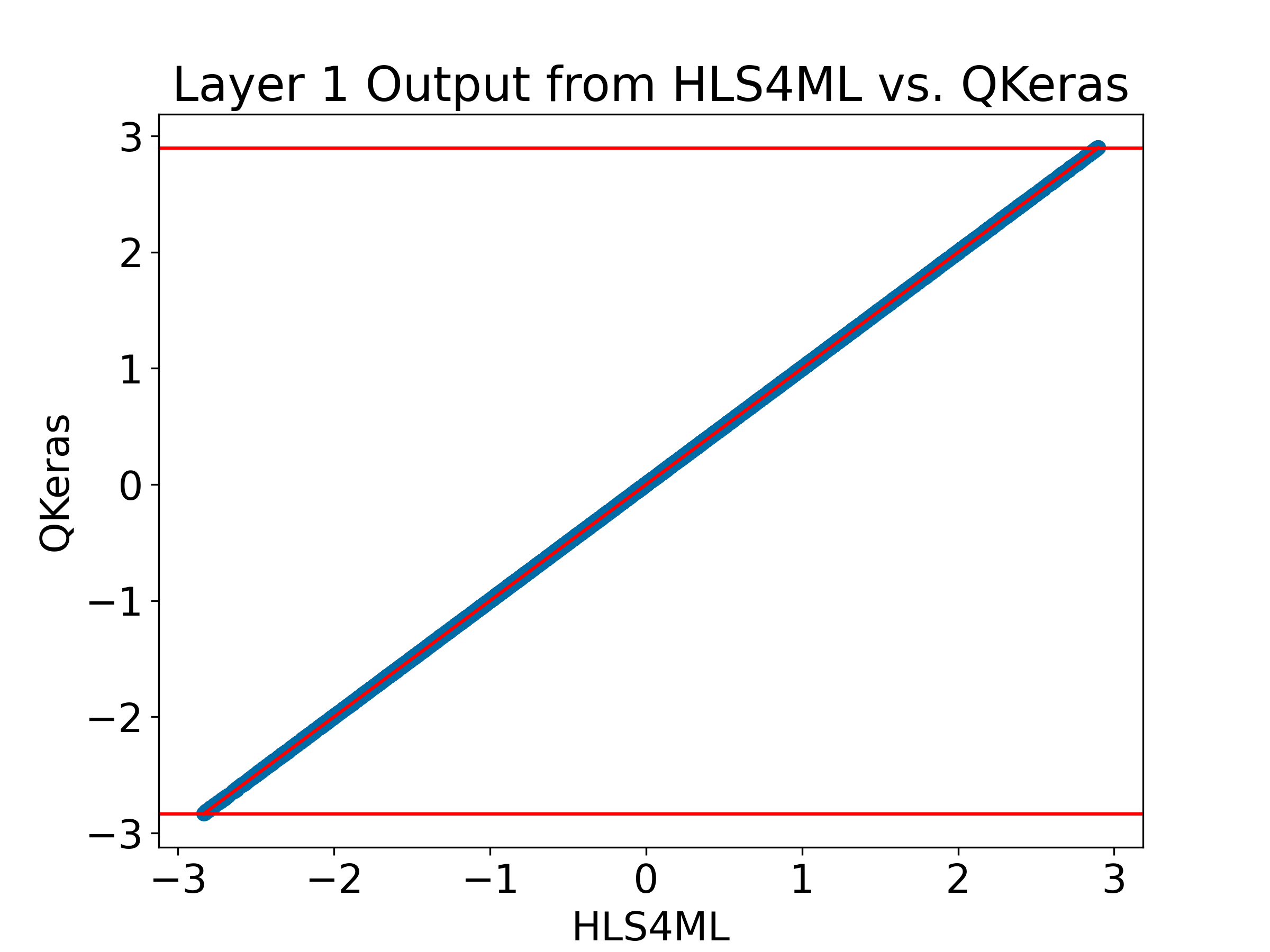}
    \caption{Trace Output with perfect result precision}
    \label{fig:trace-perfect}
\end{figure}

\begin{figure}[h]
    \centering
    \includegraphics[width=1\columnwidth]{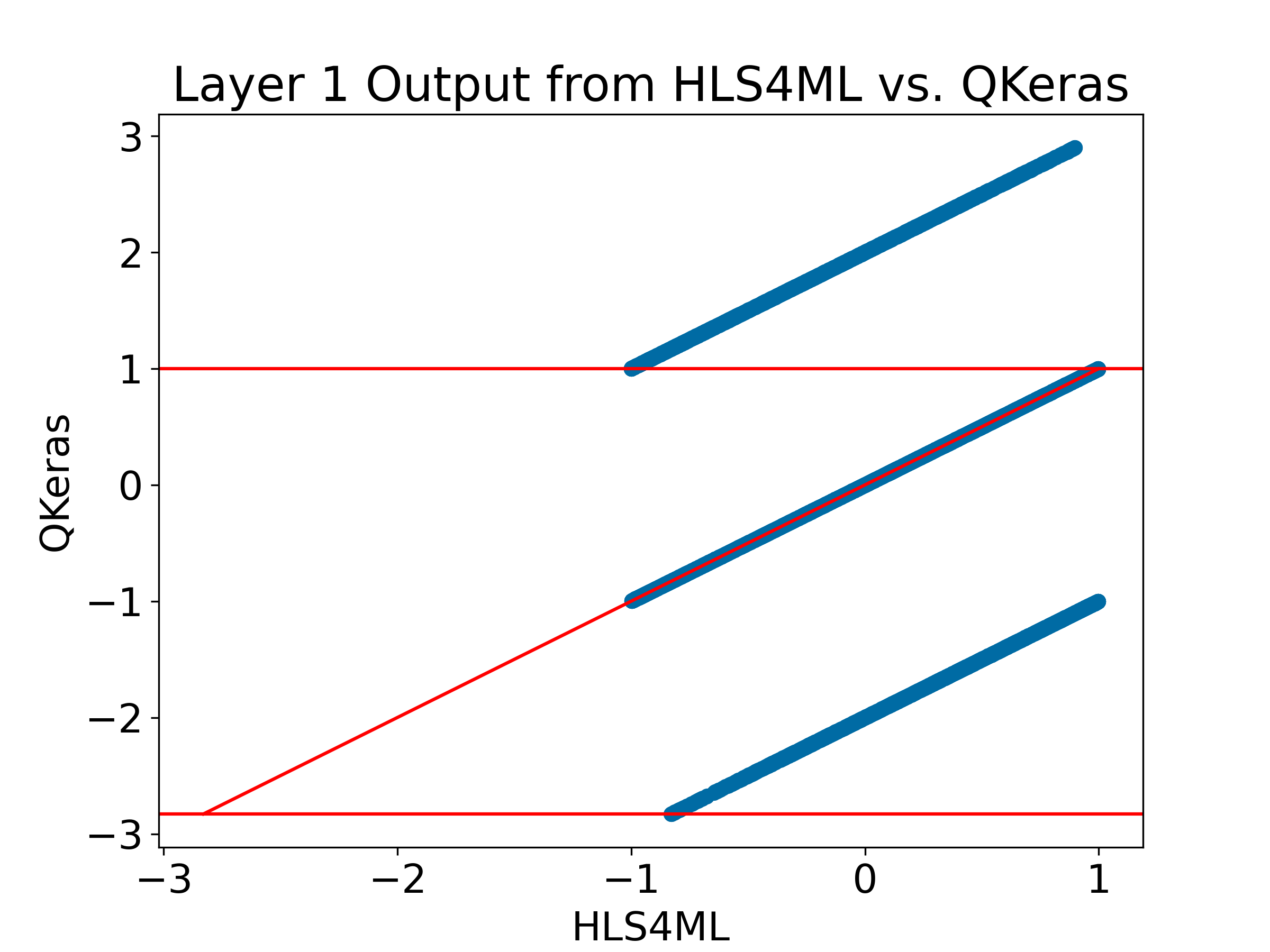}
    \caption{Trace Output with low integer result precision}
    \label{fig:trace-low-int}
\end{figure}

\begin{figure}[h]
    \centering
    \includegraphics[width=1\columnwidth]{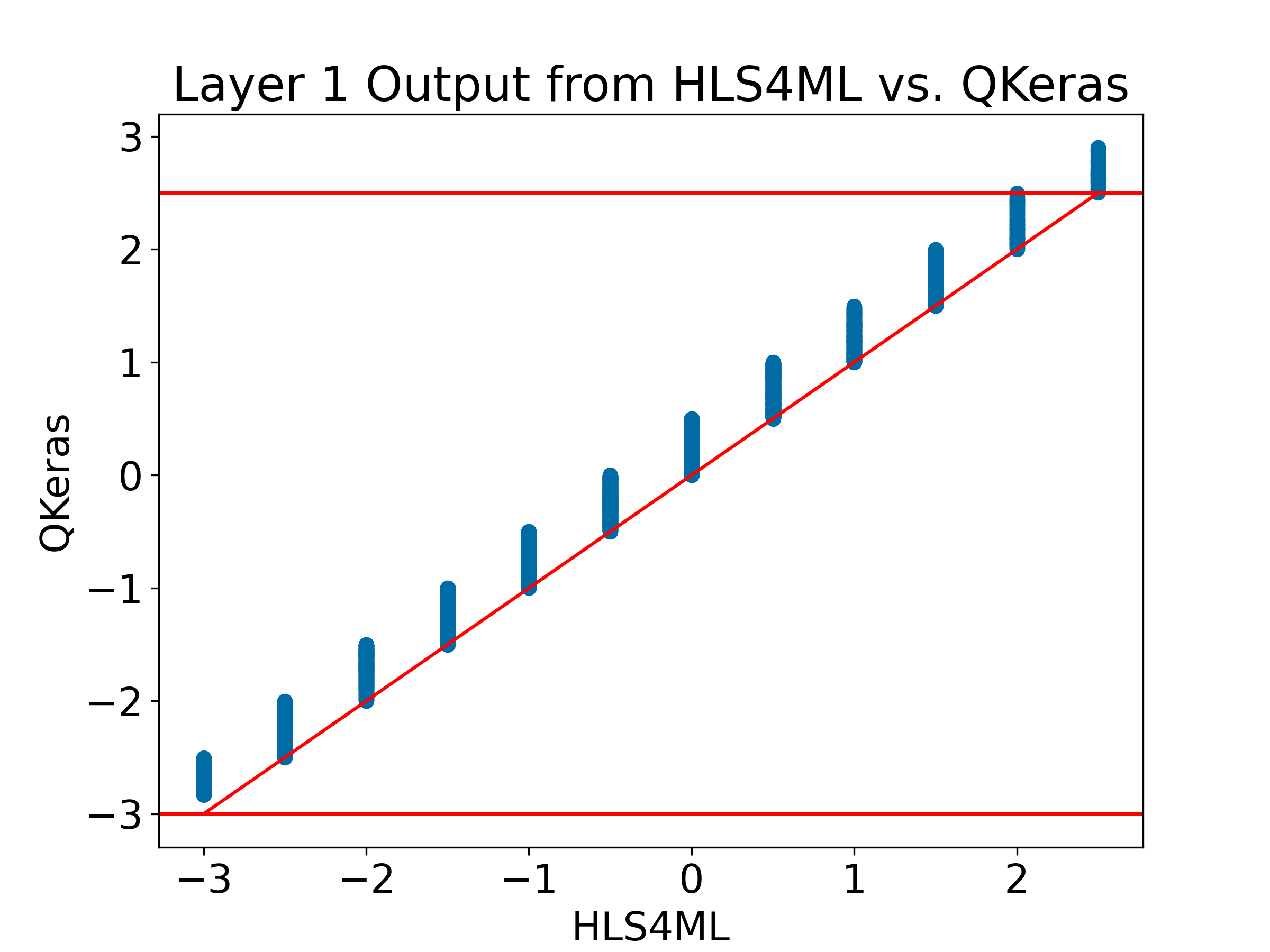}
    \caption{Trace Output with low fractional result precision}
    \label{fig:trace-low-frac}
\end{figure}

The process of fine-tuning begins with a meticulous comparison of trace outputs between the HLS4ML and QKeras models for each layer individually.
If the outputs exhibit a one-to-one linear relationship, akin to Figure \ref{fig:trace-perfect}, the result parameter's precision for that layer is deemed suitable.
Conversely, if the output comparison resembles Figure \ref{fig:trace-low-int}, indicating a disparity in integer precision (e.g., HLS4ML outputs ranging from -1 to 1 while QKeras outputs span -3 to 3), adjustments to increase integer precision become necessary.
Similarly, a pattern resembling Figure \ref{fig:trace-low-frac} suggests a need for increased fractional precision to smooth out the correlation between the two models.

\begin{figure}[h]
    \centering
    \includegraphics[width=1\columnwidth]{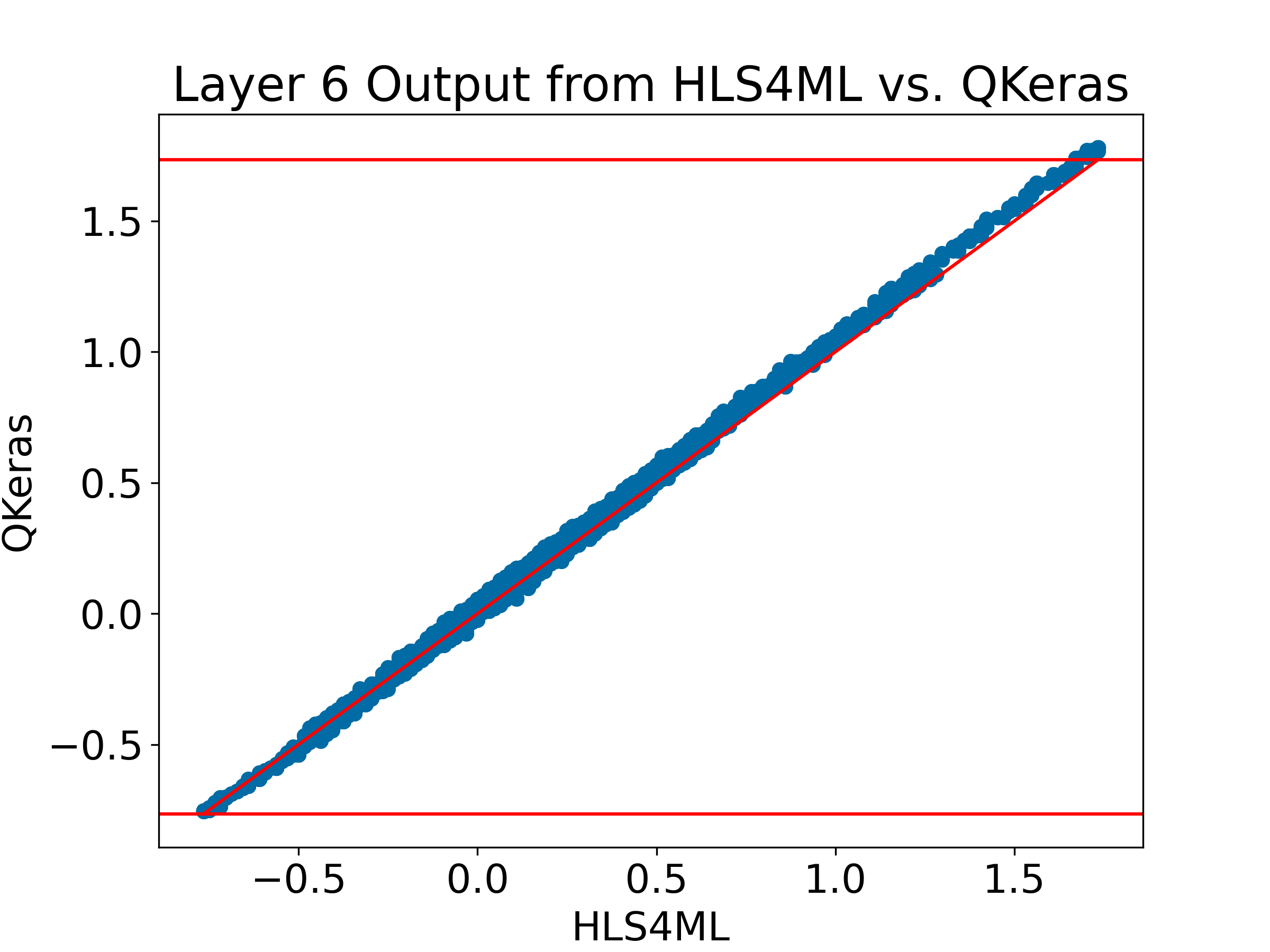}
    \caption{Trace Output for Layer 6}
    \label{fig:trace-6}
\end{figure}

\begin{figure}[h]
    \centering
    \includegraphics[width=1\columnwidth]{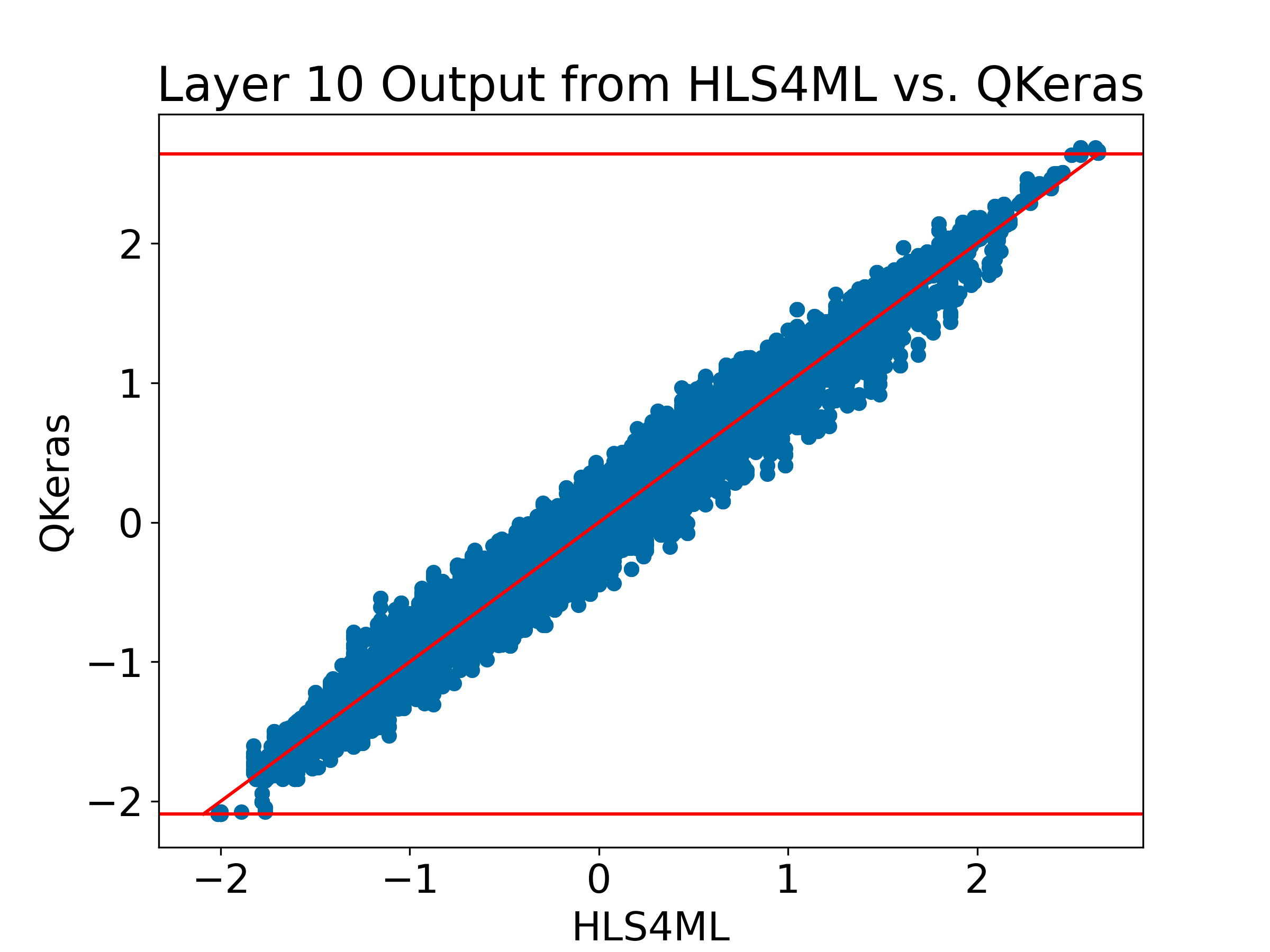}
    \caption{Trace Output for Layer 10}
    \label{fig:trace-10}
\end{figure}

Through this iterative fine-tuning process, we arrive at optimal quantization levels, as evidenced by trace graphs similarities between Figures \ref{fig:trace-perfect}, \ref{fig:trace-6} and \ref{fig:trace-10} for the 7-bit quantized weights model.
Notably, as the layer count increases, discrepancies between HLS4ML and QKeras models escalate due to precision errors propagating from one layer to the next, resulting in a cascading effect.
Nevertheless, despite these deviations, the mean squared error (MSE) loss across all HLS4ML models remains relatively low and comparable to their QKeras counterparts.

\subsection{High-Level Synthesis} 
\label{sec:hls}

\begin{table}[h!]
\centering
\begin{tabular}{|c|c|}
\hline
\textbf{Layer} & \textbf{Number of Nodes} \\
\hline
1 & 8 \\
2 & 8 \\
3 & 8 \\
4 & 8 \\
5 & 8 \\
6 & 8 \\
7 & 16 \\
8 & 16 \\
9 & 16 \\
10 & 32 \\
\hline
\end{tabular}
\caption{Number of nodes in each fully connected neural network layer}
\label{tab:nn_layers}
\end{table}

After quantization of the model, high-level synthesis is performed to reach a hardware implementation.

\subsubsection{Model Synthesis and Resource Utilization}

Once the appropriate quantization level are determined and configured in the HLS4ML model dictionary, we proceed to compile and build the model.
This results in the generation of  files.
Our target FPGA board is the Zynq UltraScale+ MPSoC ZCU102, with a target clock period set at 3.225 ns and the model strategy optimized for area.
A detailed report is then generated, outlining the achieved clock frequency and FPGA utilization percentages.
\begin{table}[h]
\centering
\begin{tabular}{|c|c|c|c|c|c|}
\hline
\textbf{Quantization}           & \textbf{Mean Infidelity } & \textbf{LUT (\%)} & \textbf{FF (\%)} \\ \hline
18      & 0.00115           & 140          & 40    \\ \hline
7& 0.00123            & 114         & 42                     \\ \hline
6 & 0.00235           & 95          & 40                 \\ \hline
5 & 0.00234          & 87         & 37                 \\ \hline
4 & 0.00919      & 69          & 23             \\ \hline
\end{tabular}
\vspace{10px}
\caption{Random\_3 model HLS results}
\label{tab:quantization-results}
\end{table}

\begin{figure}[h]
    \centering
    \includegraphics[width=1\columnwidth]{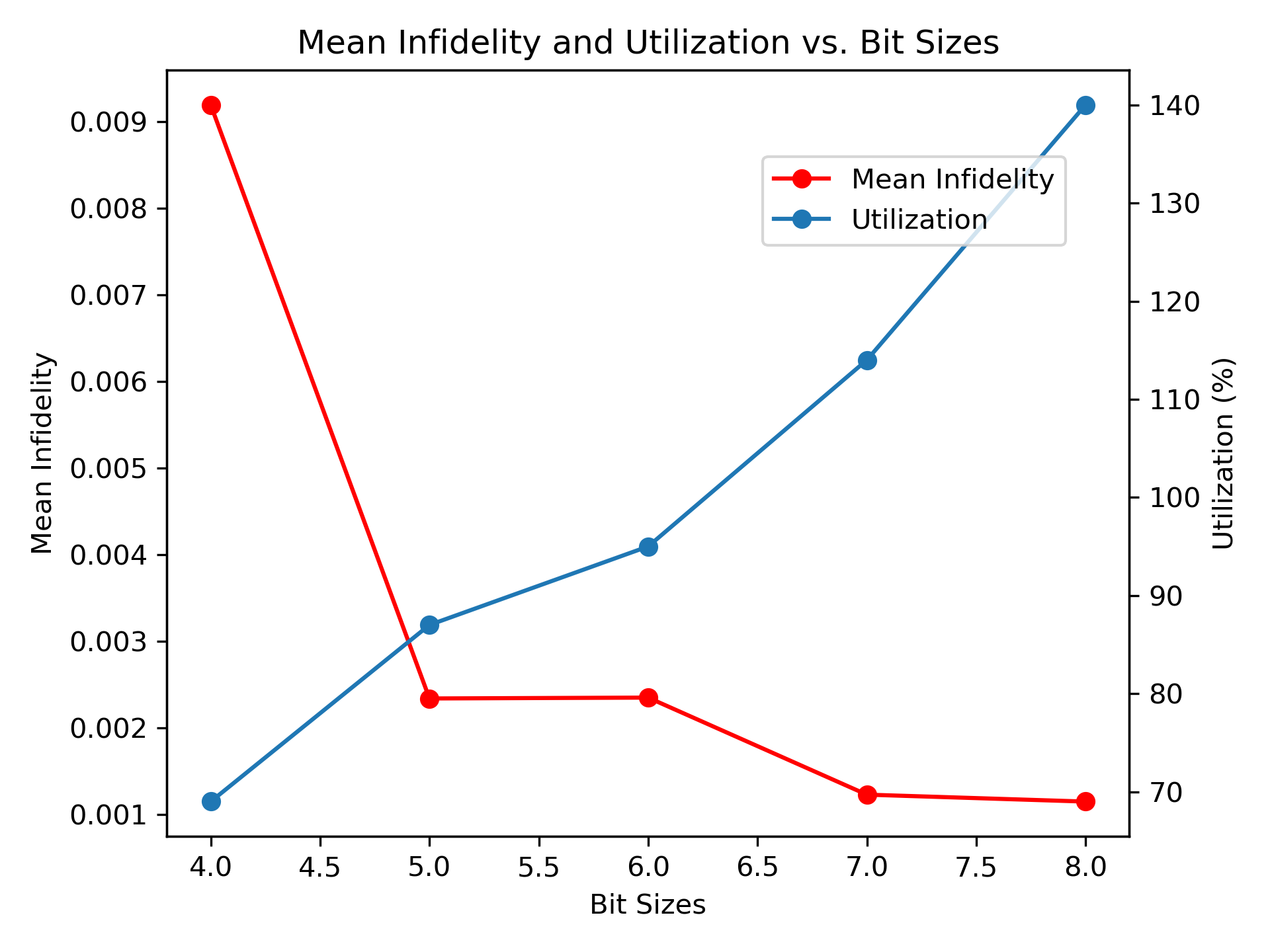}
    \caption{LUT Utilization vs. Mean Infidelity}
    \label{fig:lut-vs-fidelity}
\end{figure}

Table \ref{tab:quantization-results} presents the utilization percentages for Flip-Flops (FF) and Look-up Tables (LUT), alongside the mean infidelity achieved for each model.
For weight quantization bits 7 and 8, we observe LUT utilization exceeding 100\%, indicating these models would not be suitable for deployment on this FPGA.
However, models with weight quantization bits 6, 5, and 4 exhibit utilization below 100\% for both LUTs and FFs.
The mean infidelities achieved for bit precisions 5 and 6 are sufficiently low, approximately on the order of $10^{-3}$. 
Despite both achieving low infidelity, the model with a bit precision of 6 does not appear to be Pareto optimal compared to the model with a bit precision of 5, as the latter exhibits lower infidelity and area utilization. 
This discrepancy is evident in Fig.~\ref{fig:lut-vs-fidelity}.

\section{Conclusion}\label{sec:conclusion}

Quantum computing may offer super-polynomial asymptotic scaling advantages for some important problems, but its practical realization hinges on addressing critical challenges such the development of optimal quantum control systems with sufficiently low latency and high fidelity.
Here, we explored the fusion of machine learning and quantum control, with a focus on microwave cavity qudits.
By leveraging classical optimization for data generation and machine learning models for parameter determination, we achieved positive results in gate trace infidelity reduction and FPGA resource efficiency. 
Our findings underscore the potential of machine learning in enhancing quantum control techniques, paving the way for more robust and scalable quantum computing platforms.
As quantum technologies continue to advance, interdisciplinary approaches like the techniques presented here will play a pivotal role in harnessing their full potential and accelerating the realization of practical quantum applications.

\section{Future Work}\label{sec:future-work}

While the contributions made here are concrete and promising, optimal control for qudit devices in quantum computers remains an open research problem.
Below, we offer some specific questions that may inform future research directions.

\begin{itemize}
    \item The main issue holding back model infidelity performance without the need for further fine-tuning optimizations are the discontinuities present in the source data and the numerical stability issues related to this. 
    Using knowledge distillation, we have demonstrated that it is possible to create a data set of continuous model parameters that allows models to achieve near-Pareto performance without fine-tuning.
    It would be interesting to investigate if by applying a continuity restriction during the generation of the source data, such solutions could immediately be found.
    \item Currently, the neural networks and infidelity cost function are implemented in different frameworks, namely TensorFlow and Jax\footnote{\url{https://jax.readthedocs.io/en/latest/index.html}}.
    While integration of both functions has been achieved, this implementation prevents the joint parallel execution of both functions and thus makes the fine-tuning process more expensive.
    Implementing both in the same framework would allow for better integration and more extensive model fine-tuning.
    \item While the MLP model architectures achieve near-Pareto performance through knowledge distillation, the best performance is achieved with the MoE architectures.
    In future work, the implementation of these more complex models in hardware should be explored.
    \item In order to put the performance of the ML control model in better perspective, it should ideally be compared to the more common lookup table implementation.
    This, however, fell beyond the scope of the current project.
    \item Recent experiments use short SNAP sequences, and thus, pre-compilation of each SNAP angle by optimal control is possible \cite{quantum_rf_cavity_control_gate}. For general algorithm implementations typically involving a large number of SNAP gates \cite{kurkcuoglu2022quantumsimulationphi4theories}, a quick on-the-fly compilation using the ML control model (or a lookup table) is preferable. The comparison between the pre-compilation and on-the-fly compilation should be investigated in future work.
    \item Once models with sufficient performance are achieved and deployed in FPGA hardware, an experiment with real-time qudit control in a quantum computer should be conducted.
\end{itemize}

\pagebreak
{\small
\balance
\bibliographystyle{unsrt}
\bibliography{ref}
}





\end{document}